\documentstyle[prb,aps,multicol]{revtex}
\input epsf.tex
\begin{document} 
\draft
\title{
Thermopower of a single electron transistor in the regime of strong
inelastic cotunneling}
\author{K.~A.~Matveev$^{1}$ and A.~V.~Andreev$^{2,3}$}

\address{$^{1}$Department of Physics,  Duke University, Box 90305, 
 Durham, NC 27708\\
$^{2}$Department of Physics, University of Colorado,  CB 390, 
  Boulder, CO 80309\\
$^{3}$Bell Labs, Lucent Technologies, 600 Mountain Ave., Murray Hill, NJ
07974}

\date{January 11, 2002}
\maketitle
\begin{abstract}
  We study Coulomb blockade oscillations of thermoelectric coefficients of
  a single electron transistor based on a quantum dot strongly coupled to
  one of the leads by a quantum point contact.  At temperatures below the
  charging energy $E_C$ the transport of electrons is dominated by strong
  inelastic cotunneling.  In this regime we find analytic expressions for
  the thermopower as a function of temperature $T$ and the reflection
  amplitude $r$ in the contact.  In the case when the electron spins are
  polarized by a strong external magnetic field, the thermopower shows
  sinusoidal oscillations as a function of the gate voltage with the
  amplitude of the order of $e^{-1}|r|\frac{T}{E_C}$.  We obtain
  qualitatively different results in the absence of the magnetic field.
  At temperatures between $E_C$ and $E_C|r|^2$ the thermopower
  oscillations are sinusoidal with the amplitude of order $e^{-1}|r|^2 \ln
  \frac{E_C}{T}$.  On the other hand, at $T\ll E_C|r|^2$ we find
  non-sinusoidal oscillations of the thermopower with the amplitude $\sim
  e^{-1} |r| \sqrt{T/E_C} \ln(E_C/T)$.
\end{abstract} 
\begin{multicols}{2} 
 
\section{Introduction} 

It is well known that electric current in solids can be caused not only by
an applied electric field, but also by a temperature gradient.  This gives
rise to a number of interesting thermoelectric
phenomena.\cite{Ziman,Abrikosov} It is important to note that the
thermoelectric phenomena in metals require asymmetry between electrons and
holes.  Indeed, in a perfectly electron-hole symmetric system the
temperature gradient will cause equal in magnitude currents of electrons
and holes which result in zero net electric current.

Many recent studies of thermoelectric effects were focused on mesoscopic
systems.\cite{Anisovich,Lesovik,Spivak89,Martinis94,Beenakker92,Staring93,Dzurak97,Molenkamp98,Matveev99}
Thermoelectric properties of these systems are particularly interesting
because the electron-hole asymmetry in mesoscopic devices is usually
strong and can be controlled experimentally by tuning external parameters,
such as gate voltage or magnetic field.

Of the various thermoelectric phenomena the Peltier effect is probably the
most important for technological applications: when an electric current
$I$ is passed through a system in the absence of the temperature gradient,
it is accompanied by the heat current
\begin{equation}
  \label{eq:Peltier}
  I_Q=\Pi I.
\end{equation}
Here $\Pi$ is the Peltier coefficient.  The use of the Peltier effect has
been proposed for refrigeration in conditions when various technological
constraints, such as the size of the device, outweigh the power efficiency
considerations. The strong enhancement of the particle-hole asymmetry in
mesoscopic devices and their small size make them very promising
candidates for microrefrigerators.\cite{Martinis94}
 
In the last few years many experimental and theoretical
studies\cite{Beenakker92,Staring93,Dzurak97,Molenkamp98,Matveev99} focused
on the thermoelectric properties of single electron transistors (SET).
Thermoelectric effects in these systems can be controlled by the gate
voltage $V_g$, Fig.~\ref{fig:setup}, and exhibit characteristic Coulomb
blockade oscillations.  Most of the studies of thermoelectric effects in
the Coulomb blockade regime concentrated on the thermopower
\begin{equation}
\label{eq:S}
S=-\left.\frac{V}{\Delta T}\right|_{I=0}.
\end{equation}
Here $V$ is the voltage induced across the device in the absence of net
electric current when the temperatures of the two leads differ by ${\Delta
  T}$, Fig.~\ref{fig:setup}(a).  The Peltier coefficient $\Pi$ is related
to the thermopower $S$ by an Onsager relation $\Pi=ST$.  

The theory of the Coulomb blockade oscillations in the thermopower of
single electron transistors in the weak tunneling regime was developed in
Ref.~\onlinecite{Beenakker92}.  This theory takes into account only the
lowest order tunneling processes, i.e. the sequential tunneling, and
neglects the cotunneling processes.  Its results were in agreement with
the experiments of Ref.~\onlinecite{Staring93}.  Later\cite{Dzurak97} it
became possible to experimentally access the regime of lower temperatures
and stronger coupling to the leads, where the cotunneling processes become
dominant.  The theoretical description of this regime was recently given
in Ref.~\onlinecite{Matveev99}.

In experiments with GaAs heterostructures the quantum dot is connected to
the leads by quantum point contacts.  Each contact is usually in the
regime when only one transverse mode can propagate through it, and the
transmission coefficient for this mode can be controlled experimentally.
Recently\cite{Molenkamp98} the Coulomb blockade oscillations in the
thermopower of a SET with the quantum dot strongly coupled to one of the
leads were studied for various values of the reflection coefficient
$|r|^2$ in this contact.  The setup of these experiments is schematically
shown in Fig.~\ref{fig:setup}(b).  In the regime of strong coupling,
$|r|^2\ll1$, nearly sinusoidal oscillations of the thermopower as a
function of the gate voltage $V_g$ were observed.

\narrowtext{ 
\begin{figure} 
\begin{center} 
\epsfxsize=8cm 
\epsfbox{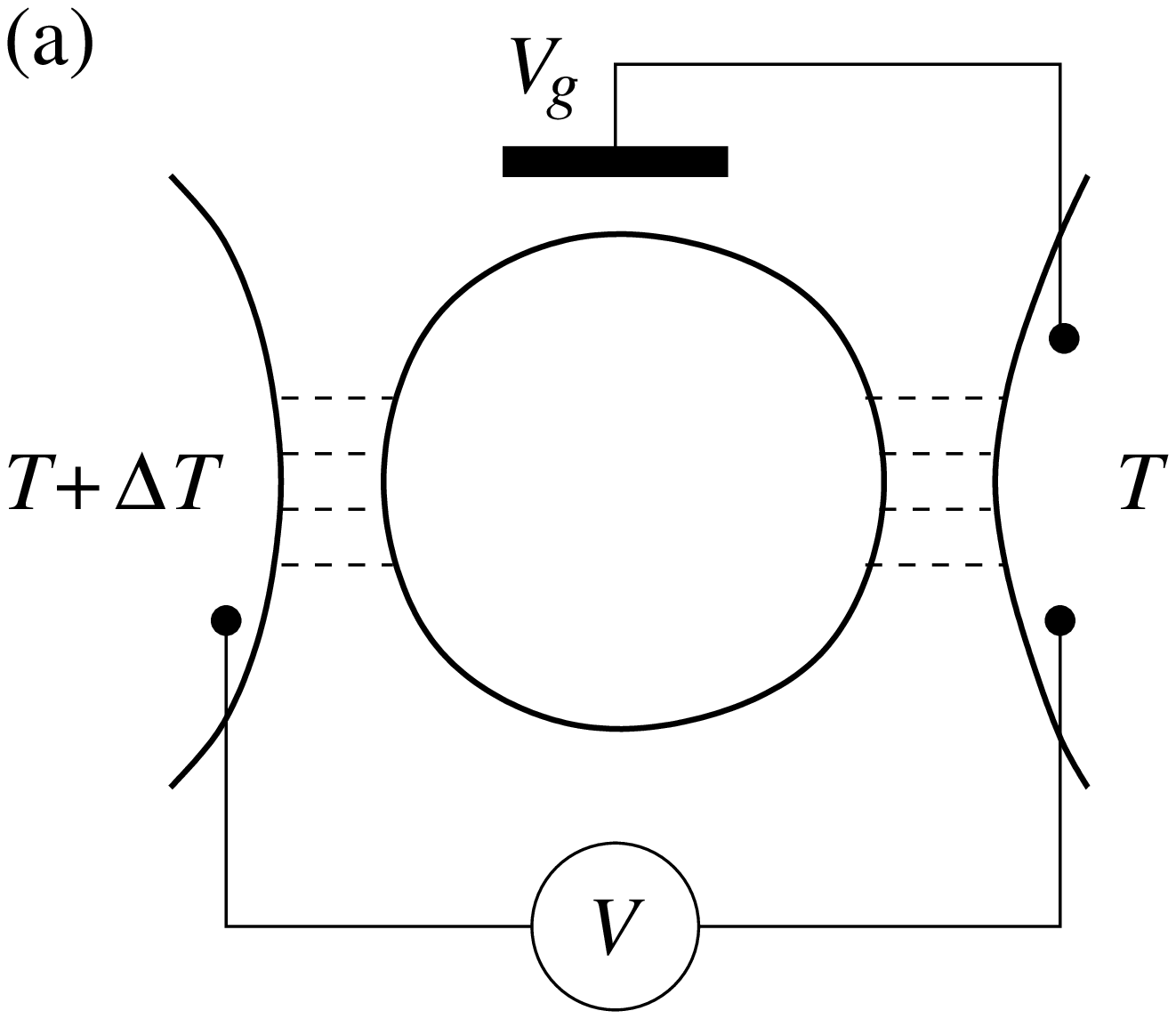}\nopagebreak\\
\epsfxsize=8cm 
\epsfbox{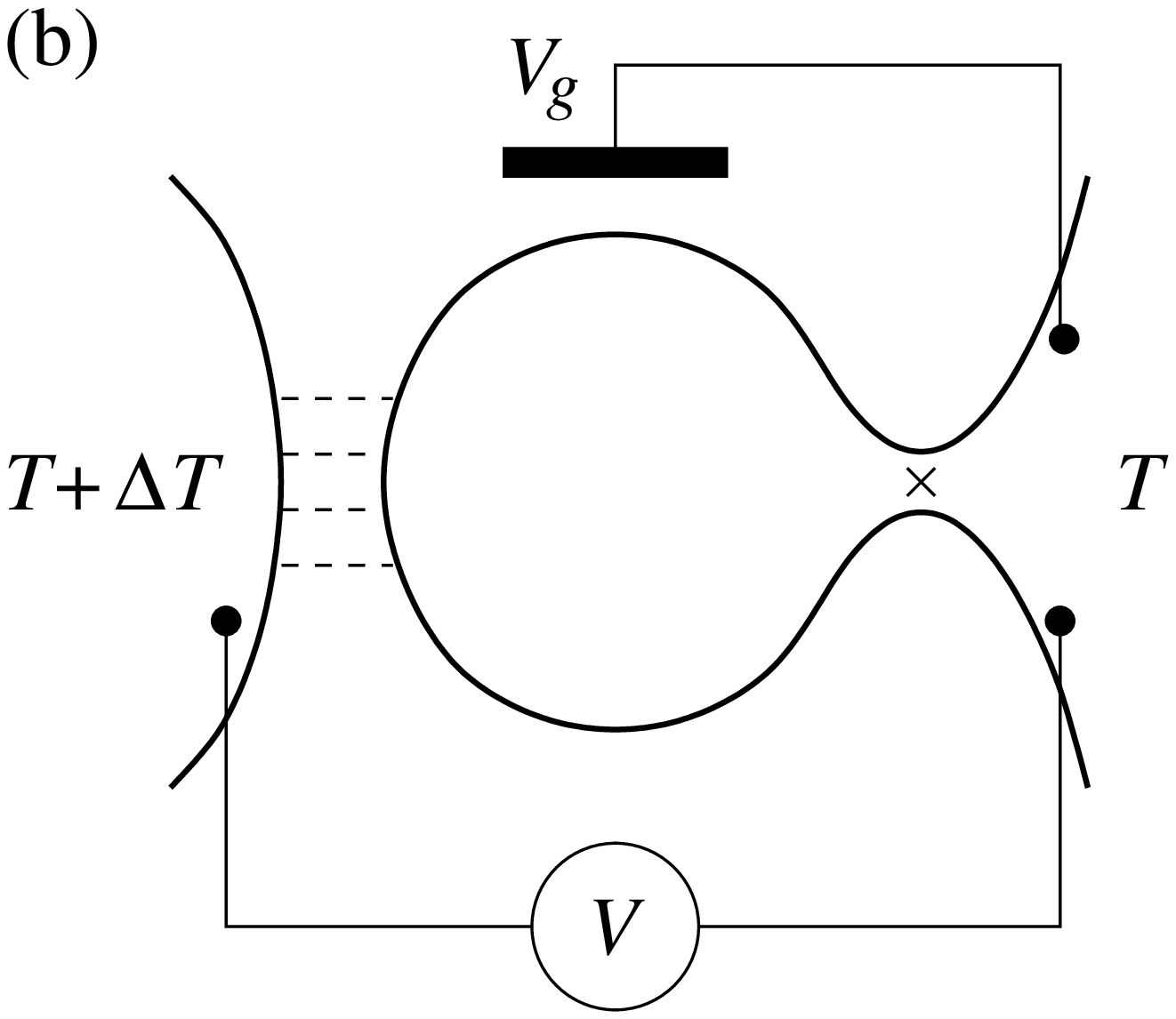}
\end{center} 
\caption{(a) Setup of the thermopower measurement in a single electron 
  transistor. A quantum dot is capacitively coupled to the gate and
  connected to the two leads by tunneling junctions. The left and right
  leads are maintained at temperatures $T+\Delta T$ and $T$, respectively,
  and the voltage $V$ across the device is measured. The thermopower
  (\protect\ref{eq:S}) is measured as a function of the gate voltage
  $V_g$. (b) A SET with a quantum dot strongly coupled to the right lead
  by a single-channel quantum point contact.  The cross $\times$ in the
  constriction represents the backscattering in the contact resulting in a
  finite reflection coefficient $|r|^2$.}
\label{fig:setup}
\end{figure} 
}

The theories of thermopower for the weak tunneling regime developed in
Refs.~\onlinecite{Beenakker92,Matveev99} rely upon the perturbation theory
in the strength of coupling between the quantum dot and the leads.  This
perturbative approach fails when the coupling to the leads is strong.  The
previous theoretical work on Coulomb blockade systems with strongly
coupled quantum dots\cite{Flensberg93,Matveev95,Furusaki95,Aleiner97} was
devoted to the studies of their thermodynamic properties and conductance.
These properties are not sensitive to the electron-hole asymmetry, and the
calculation of the thermoelectric properties requires a non-trivial
generalization of the approach of Refs.~\onlinecite{Furusaki95,Aleiner97}. 

In this paper we develop a theory of the thermopower in a SET with a
quantum dot strongly coupled to one of the leads, Fig.~\ref{fig:setup}(b).
We consider a relatively large quantum dot in which the quantum level
spacing is small, and the electron transport is dominated by inelastic
cotunneling.  In Sec.~\ref{sec:qualitative} we present a qualitative
discussion of thermoelectric transport in a SET.  The thermopower of a SET
in the regime of strong inelastic cotunneling is developed in
Sec.~\ref{sec:main} for the simpler case of spin-polarized electrons as
well as for the more interesting spin-degenerate case.  We discuss the
results and compare them with experiments in Sec.~\ref{sec:discussion}.

\section{Qualitative discussion of the thermopower at weak inelastic  
cotunneling} 
\label{sec:qualitative}

The physical meaning of the thermopower can be deduced from the Onsager
relation $S=\Pi/T$.  One can easily see from Eq.~(\ref{eq:Peltier}) that
the Peltier coefficient is determined by the average energy
$\langle\epsilon\rangle$ of the electrons carrying current through the
system, measured from the chemical potential:
$\Pi=-\langle\epsilon\rangle/e$.  Thus the thermopower measures the
average energy of the tunneling electrons in units of the temperature:
\begin{equation}
  \label{eq:meaning}
  S=-\frac{\langle\epsilon\rangle}{eT}.
\end{equation}
Here $e$ is the absolute value of the electron charge.  The average energy
of the charge carriers $\langle\epsilon\rangle$ is determined by a
particular mechanism of transport through the system. 

A conventional SET schematically shown in Fig.~\ref{fig:setup}(a) consists
of a quantum dot weakly coupled to two leads.  The transport of electrons
from the left lead to the right one is achieved by either sequential
tunneling or cotunneling (elastic or inelastic).  

The sequential tunneling refers to the lowest-order tunneling processes in
which one electron tunnels into or out of the dot.  As a result of each
tunneling event the charge of the dot changes by $\pm e$.  When an
electron tunnels into or out of the dot, the electrostatic energy of the
system increases by $u_+$ or $u_-$, respectively.  The values of $u_+$ and
$u_-$ are of the order of the charging energy $E_C=e^2/2C$, where $C$ is
the capacitance of the dot; their values can be tuned by adjusting the
gate voltage $V_g$.  The electron that tunnels into the dot has to have
the energy $\epsilon\approx u_+$ in order to charge the dot, so the
tunneling is exponentially suppressed at low temperatures as $e^{-u_+/T}$.
Similarly the rate of tunneling out of the dot is suppressed as
$e^{-u_-/T}$.  The thermopower of a SET in the regime of sequential
tunneling was studied in Ref.~\onlinecite{Beenakker92}.

The cotunneling mechanism accounts for the coherent second-order tunneling
processes in which at the first step an electron tunnels from, say, the
left lead into the dot, and at the second step this or another electron
(for elastic and inelastic cotunneling, respectively) tunnels from the dot
to the right lead.  Since as a result of cotunneling processes the charge
of the dot remains unchanged, the electrons participating in the transport
through the dot do not need to have large energies $\sim u_\pm$, and the
transport is not suppressed exponentially at low temperatures.  It is
important to note that the elastic cotunneling involves elastic
propagation of electrons between the tunneling contacts in the dot.  The
resulting contribution to the transport is inversely proportional to the
volume of the dot.  Thus in relatively large dots the finite temperature
transport is dominated by inelastic cotunneling processes.

In order to find the cotunneling thermopower of a SET, one needs to
evaluate the average energy $\langle\epsilon\rangle$ of the electron
transferred through the device.  Inelastic cotunneling involves electrons
within the energy strip of width $\sim T$ near the Fermi level.  The
probability $w(\epsilon)$ of the second-order tunneling process is
inversely proportional to the square of the energy of the virtual state,
\begin{equation}
  \label{eq:prob}
  w(\epsilon)\propto\left(\frac{1}{u_+ + \epsilon' - \epsilon}
                    +\frac{1}{u_- + \epsilon - \epsilon'}\right)^2,
\end{equation}
where we assumed that the electron with energy $\epsilon$ tunnels into a
state of  energy $\epsilon'$ in the dot.  The energies $\epsilon$ and
$\epsilon'$ are of the order of $T$ and small compared to the charging
energies $u_\pm$.  One can therefore expand Eq.~(\ref{eq:prob}) in small
$(\epsilon-\epsilon')/u_\pm$, 
\begin{equation}
  \label{eq:prob2}
  w(\epsilon)\propto\left(\frac{1}{u_+}+\frac{1}{u_-}\right)^2 
               \left[1+2\left(\frac{1}{u_+}
                      -\frac{1}{u_-}\right)(\epsilon-\epsilon')
               \right].
\end{equation}

The $\epsilon$-dependent correction in Eq.~(\ref{eq:prob2}) shows that the
tunneling probability increases or decreases with the energy $\epsilon$ of
the tunneling particle depending on the sign of $(u_+^{-1}-u_-^{-1})$.
Since the typical energy $\epsilon\sim T$, the relative magnitude of the
term breaking the electron-hole symmetry in (\ref{eq:prob2}) is $\sim
T(u_+^{-1}-u_-^{-1})$ and the average energy of the tunneling particles is
$\langle \epsilon \rangle \sim T^2 (u_+^{-1}-u_-^{-1})$. Then using
Eq.~(\ref{eq:meaning}) we estimate the thermopower of a SET as
\begin{equation}
  \label{eq:S_weak}
  S=\lambda\frac{T}{e}\left(\frac{1}{u_-}-\frac{1}{u_+}\right).
\end{equation}
Here $\lambda$ is a numerical coefficient of order unity; its value
$\lambda=4\pi^2/5$ has been found in Ref.~\onlinecite{Matveev99}.

An interesting feature of the result (\ref{eq:S_weak}) is that the
thermopower does not depend on the strength of coupling of the dot to the
leads.  This feature is expected to persist as long as the transmission
coefficients of the barriers are small.  However, as the barrier
approaches the regime of perfect transmission, the Coulomb blockade
oscillations of physical quantities are expected to
disappear.\cite{Matveev95,Furusaki95}  One should therefore expect that in
the regime of strong coupling the thermopower will depend on the
reflection coefficient $|r|^2$ of the contact.

It is also worth mentioning that the thermopower $S\sim e^{-1} T/E_C$ in a
SET is much greater than the typical value $S\sim e^{-1} T/E_F$ of the
thermopower in metals; here $E_F$ is the Fermi energy.  This is a
consequence of the fact that the charging effects in the dot enhance the
electron-hole asymmetry.  We will show in Sec.~\ref{sec:main} that such
behavior of the thermopower at $T\to0$ persists in the strong coupling
regime.

\section{Thermopower in the regime of strong inelastic cotunneling}
\label{sec:main}

In this section we calculate the thermopower $S$ of a SET in which one of
the contacts is in the strong tunneling regime, Fig.~\ref{fig:setup}(b).
We restrict ourselves to the linear response regime when both the
temperature difference between the leads $\Delta T$ and the voltage $V$
across the device are small.  The thermopower (\ref{eq:S}) is defined in
terms of the voltage $V$ induced across the device by the temperature
difference $\Delta T$ at zero current. In practice, however, it is easier
to calculate the current response
\begin{equation}
I = G V +  G_T \Delta T.
\label{eq:linresp}
\end{equation}
Here $G$ is the conductance of the SET, and $G_T$ is the thermoelectric
coefficient describing the current response to an applied temperature
difference.  The thermopower (\ref{eq:S}) can then be expressed as
\begin{equation} 
\label{eq:thermopowerdef} 
S=\frac{G_T}{G}.
\end{equation} 
The conductance $G$ of a SET in the regime of strong inelastic cotunneling
was found in Ref.~\onlinecite{Furusaki95}.  Hence in the following we
concentrate on the calculation of the thermoelectric coefficient $G_T$.

\subsection{Tunneling approximation}
\label{sec:tunneling}

We assume that the conductance of the tunneling junction connecting the
dot to the left lead is much smaller than the conductance quantum, $G_L
\ll e^2/h$.  In this case one can describe the SET by the Hamiltonian
$H=H_L + H'$, where $H_L$ is the Hamiltonian describing the tunneling of
electrons in the left contact,
\begin{eqnarray}
  \label{eq:H_L}
  H_L&=&\sum_{k}\epsilon_k a^\dagger_{k} a_{k} 
      +\sum_{p}\epsilon_p a^\dagger_{p} a_{p}\nonumber\\
      &&+\sum_{kp} (t_{kp} a^\dagger_{k}a_{p}
        +t^*_{kp} a^\dagger_{p}a_{k}),
\end{eqnarray}
and $H'$ accounts for the transport through the right junction and the
electron-electron interactions in the system.  Here $a_k$ and $a_p$ are
the electron annihilation operators in the left lead and in the dot,
respectively, $\epsilon_k$ and $\epsilon_p$ are the energies of the
corresponding states, the matrix elements $t_{kp}$ describe the weak
tunneling of the electrons through the barrier.

We will account for the tunneling through the left barrier in the lowest
(second) order of the perturbation theory in $t_{kp}$.  In this
approximation all of the temperature drop occurs at the tunneling barrier.
We take the temperature of the left lead to be $T+\Delta T$ and that of
the dot to be $T$.  Let us denote the tunneling density of states in the
left lead by $\nu_l(\epsilon)$ and that in the dot by $\nu(\epsilon)$.  To
the lowest order in the tunneling matrix element the current through the
tunneling contact can be obtained with the aid of the Fermi golden rule:
\begin{equation}
\label{eq:A1}
I=-2\pi e \langle |t_{kp}|^2 \rangle  
   \int^{\infty}_{-\infty} \nu_l(\epsilon)\nu (\epsilon)
   \left[n_l(\epsilon)-n (\epsilon)\right] d\epsilon.
\end{equation}
Here $n_l(\epsilon)$ and $n(\epsilon)$ denote the Fermi distribution
functions at the temperature of the left lead $T+\Delta T$ and the dot
$T$, respectively.  The square of the tunneling matrix element in
Eq.~(\ref{eq:A1}) is averaged over the states near the Fermi level.

To determine $G_T$ we assume that the chemical potentials in the Fermi
functions in Eq.~(\ref{eq:A1}) are the same and expand the current to
first order in $\Delta T$,
\begin{equation}
\label{eq:A2}
I=-2\pi e \nu_l \langle |t_{kp}|^2 \rangle  \frac{\Delta T}{4 T^2}
   \int^{\infty}_{-\infty}
    \frac{\epsilon \nu (\epsilon)}{\cosh^2 \left(\frac{\epsilon}{2 T}\right)}
    d\epsilon.
\end{equation}
Here we have replaced the density of states in the left contact
$\nu_l(\epsilon)$ by its value $\nu_l$ at the Fermi energy using its weak
energy dependence.  The corrections to this approximation are small in the
ratio of the temperature to the Fermi energy.  We will see below that the
density of states in the dot has the energy dependence at the much smaller
energy scale $T$ due to the electron-electron interactions.

In order to express the thermoelectric coefficient $G_T$ in terms of
physically measurable quantities, we use Eq.~(\ref{eq:A1}) to calculate
the conductance $G_L$ of the left barrier assuming that the electrons in
the dot are non-interacting.  This conductance can, in principle, be
measured experimentally by opening completely the constriction connecting
the dot to the right lead.  The result is expressed in terms of the
density of states in the dot $\nu_0$, which is no longer renormalized by
the electron-electron interactions:
\begin{equation}
\label{eq:A3}
G_L=2\pi e^2  \nu_l \nu_0 \langle |t_{kp}|^2 \rangle.
\end{equation}
Using Eqs.~(\ref{eq:A3}) and (\ref{eq:A2}) we can now express
$G_T=I/\Delta T$ as
\begin{equation} 
  \label{eq:thermocurrent} 
  G_T=-\frac{G_L}{4 T^2 e \nu_0}\int_{-\infty}^\infty 
  \frac{\epsilon \nu(\epsilon)}{\cosh^2 \left( \frac{
        \epsilon}{2T}\right)}  d\epsilon.
\end{equation} 
 
Equation (\ref{eq:thermocurrent}) reduces our problem to the calculation
of the energy-dependent tunneling density of states $\nu(\epsilon)$.  We
note that unlike the conductance $G$ of the SET, the thermoelectric
coefficient $G_T$ is determined by the odd (in energy) component of
density of states $\nu(\epsilon)$. Therefore the thermopower measurements
represent an independent test of the theory of Coulomb blockade in nearly
open dots developed in Refs.~\onlinecite{Matveev95,Furusaki95,Aleiner97}.

It is well known that the tunneling density of states can be expressed
in terms of the electron Green's function.  A specific form of this
relation that will be convenient for further calculations is
\begin{equation} 
  \label{eq:dos} 
  \nu(\epsilon)=-\frac{1}{\pi} \cosh \frac{\epsilon}{2T} 
  \int_{-\infty}^{\infty} {\cal G}\left( \frac{1}{2T}+i t\right) 
  \exp(i\epsilon t) \, dt.
\end{equation} 
Here ${\cal G}(\tau)=-\langle T_\tau\psi_L(\tau)\psi_L^\dagger(0)\rangle$
is the Matsubara Green's function; $\psi_L$ is the annihilation operator
of an electron in the dot at the position of the left contact.  For the
derivation of Eq.~(\ref{eq:dos}) see Appendix~\ref{sec:dos}.  Substituting
Eq.~(\ref{eq:dos}) into Eq.~(\ref{eq:thermocurrent}), we express the
thermoelectric coefficient $G_T$ in terms of the Green's function:
\begin{equation} 
  \label{eq:thermocond} 
  G_T=\frac{i\pi G_L}{2e\nu_0} \int_{-\infty}^\infty 
       \frac{\sinh(\pi T t)}{\cosh^2 (\pi T t)}\,
       {\cal G}\left(\frac{1}{2T}+it\right) dt.
\end{equation} 
This expression is insensitive to the specific form of the
interactions in the quantum dot and its coupling to the right lead.
In the following sections we calculate the Green's function ${\cal
  G}(\tau)$ in the strong inelastic cotunneling approximation and find
the corresponding value of $G_T$.

\subsection{Inelastic cotunneling approximation}

At finite temperature in a sufficiently large dot one can use the
approximation of inelastic cotunneling which neglects the possibility
of elastic propagation of electrons between the two contacts in the dot.
In this case the system can be modeled by the Hamiltonian 
\begin{equation}
  \label{eq:hamiltonian}
  H=H_L+H_R+H_C,
\end{equation}
in which the Hamiltonians $H_L$ and $H_R$ describe the independent
subsystems of electrons propagating through the left and right contacts;
in particular $[H_L,H_R]=0$.  We have discussed the form of $H_L$ in
section~\ref{sec:tunneling}, see Eq.~(\ref{eq:H_L}).  For weak inelastic
cotunneling $H_R$ would have a form similar to Eq.~(\ref{eq:H_L}).  In the
case of strong inelastic cotunneling the Hamiltonian $H_R$ has a
completely different form which will be discussed in
sections~\ref{sec:spinless} and \ref{sec:spinful}.  Finally, the term
$H_C$ describes the Coulomb interactions in the dot.  At low energies the
interactions are adequately accounted for by the charging energy
approximation:
\begin{equation}
  \label{eq:oldH_C}
   H_C = E_C( \hat{n}_L + \hat{n}_R -N )^2.
\end{equation}
Here $\hat n_L$ and $\hat n_R$ are the operators of the number of
electrons that entered the dot through the left and right contacts,
respectively, and $N$ is a dimensionless parameter which is proportional
to the gate voltage $V_g$.  

By definition of the operator $\hat n_L$ its commutation relations with
the fermion operator $\psi_L$ defined below Eq.~(\ref{eq:dos}) have the
form $[\psi_L,\hat n_L]=\psi_L$.  For the convenience of the following
calculations we will rewrite the charging energy (\ref{eq:oldH_C}) in the
form
\begin{equation}
  \label{eq:H_C}
   H_C = E_C( \hat{n} + \hat{n}_R -N )^2,
\end{equation}
where $\hat n$ is an integer-valued operator that commutes with $\psi_L$.
In order to preserve the commutation relations between $\psi_L$ and $H_C$
we replace $\psi_L\to\psi_L F$, where $F$ is the operator lowering $\hat
n$ by unity: $[F,\hat n]=F$.  It is important to note that the
substitution $\psi_L\to\psi_L F$ does not affect the form of the
Hamiltonian $H_L$, and the only modification in the discussion of
Sec.~\ref{sec:tunneling} is in the definition of the Green's function in
Eq.~(\ref{eq:dos}),
\begin{equation}
  \label{eq:G}
  {\cal G}(\tau) = 
    -\langle T_\tau\psi_L(\tau)F(\tau)F^\dagger(0)\psi_L^\dagger(0)\rangle.
\end{equation}
The operators $\psi_L$ and $\psi_L^\dagger$ now commute with $H_R+H_C$,
whereas $F$ and $F^\dagger$ commute with $H_L$.  Consequently, the Green's
function (\ref{eq:G}) factorizes,
\[
    {\cal G}(\tau) = 
    -\langle T_\tau\psi_L(\tau)\psi_L^\dagger(0)\rangle
     \langle T_\tau F(\tau)F^\dagger(0)\rangle.
\]
In the new representation the operators $\psi_L$ and $\psi_L^\dagger$
describe non-interacting fermions, whose Green's function is well known.
We can then rewrite ${\cal G}(\tau)$ as
\begin{eqnarray} 
\label{eq:factorization} 
    {\cal G}(\tau)&=&-\frac{\nu_0 \pi T}{\sin(\pi T \tau)}K(\tau),\\
\label{eq:kdef}
    K(\tau)&=&\langle T_\tau F(\tau)F^\dagger(0)\rangle. 
\end{eqnarray} 

Substituting Eq.~(\ref{eq:factorization}) into Eq.~(\ref{eq:thermocond})
we express the thermoelectric coefficient $G_T$ of the dot in terms of the
correlator $K(\tau)$,
\begin{equation} 
\label{eq:thermocondk} 
  G_T=-\frac{i\pi^2}{2} \frac{G_LT}{e}
       \int_{-\infty}^\infty \frac{\sinh(\pi Tt)}{\cosh^3(\pi Tt)} 
         K\left(\frac{1}{2T}+it\right) dt.
\end{equation} 
Unlike Eq.~(\ref{eq:thermocond}), formula (\ref{eq:thermocondk}) assumes
that the transport through the dot is due to the inelastic cotunneling
mechanism, and that the electron-electron interactions in the system are
completely described by the charging energy (\ref{eq:H_C}).  On the other
hand, the coupling $H_R$ of the dot to the right lead is still arbitrary.
In particular, Eq.~(\ref{eq:thermocondk}) is valid in the case of a
metallic grain coupled to the lead by a wide contact supporting many
channels.

In this paper we consider the case of a single-channel contact, which is
usually realized in semiconductor devices.  Depending on the presence of
a magnetic field polarizing the spins of the electrons, one has to
consider the cases of either spinless or spin-$\frac12$ electrons.

\subsection{Spinless electrons}
\label{sec:spinless}

We start with the simpler case of spinless electrons.  Following
Ref.~\onlinecite{Matveev95,Furusaki95} we describe the electron transport
through the right quantum point contact by a model of one-dimensional
fermions.  In the case of strong coupling of the dot to the right lead,
the charging energy (\ref{eq:H_C}) gives rise to non-trivial Coulomb
correlations of the motion of electrons through the constriction.  It is
more convenient to treat the problems of interacting one-dimensional
electrons in the bosonized representation.  Then the
Hamiltonian\cite{Matveev95,Furusaki95} of the right constriction takes the
form $H_R=H_R^{(0)} + H_R'$, where
\begin{mathletters}
  \label{eq:spinless_hamiltonian}
\begin{eqnarray}
  \label{eq:freehamiltonian} 
  H_R^{(0)}&=&\frac{v_F}{2\pi} \int_{-\infty}^\infty 
             \left\{\pi^2\Pi^2(x)+[\partial_x\phi(x)]^2\right\}dx ,\\ 
  \label{eq:scatteringspinless}
  H_R' &=&  -\frac{D}{\pi}|r|\cos[2\phi(0)].
\end{eqnarray} 
Here $\phi$ and $\Pi$ are bosonic fields satisfying the commutation
relations $[\phi(x),\Pi(y)]=i\delta(x-y)$, the parameter $v_F$ is the
Fermi velocity of the electrons, $r$ is the reflection amplitude in the
constriction, and $D$ is the bandwidth.  The regions $x<0$ and $x>0$ in
the integral of Eq.~(\ref{eq:freehamiltonian}) represent the electrons in
the dot and in the right lead, respectively.  The deviation of the density
of one-dimensional electrons from its ground state value is given by
$\partial_x\phi(x)/\pi$.  Thus the number of electrons that have entered the
dot through the right constriction is $n_R=\phi(0)/\pi$, and the charging
energy (\ref{eq:H_C}) takes the form
\begin{equation}
  \label{eq:H_Cspinless}
   H_C = E_C \left[\hat{n} + \frac{1}{\pi} \phi(0) - N \right]^2.
\end{equation}
\end{mathletters}
The advantage of the bosonization approach is that the Coulomb interaction
term (\ref{eq:H_Cspinless}) is quadratic in the bosonic operator $\phi$,
and, therefore, can be treated exactly.  On the other hand, the
backscattering of electrons in the constriction in the bosonized
representation takes the strongly non-linear form
(\ref{eq:scatteringspinless}).  As a result the backscattering can only be
accounted for perturbatively, using the small parameter $|r|<1$.

We will calculate the time-ordered correlator $K(\tau)$ defined by
Eq.~(\ref{eq:kdef}) as an imaginary-time functional integral over the
bosonic field $\phi$.  The operator $F^\dagger(0)$ increases $n$ from 0 to
1 at time $t=0$, whereas $F(\tau)$ changes it back to $n=0$ at time
$t=\tau$.  Therefore $F(0)F^\dagger(0)$ in the functional integral can be
omitted provided that the operator $\hat n$ in the action is replaced by
\begin{equation} 
  \label{eq:ntau} 
  n_\tau (t)=\theta(t) \theta(\tau -t).
\end{equation} 
Here $\theta(t)$ is the unit step function.  Upon this procedure the
correlator $K(\tau)$ is expressed as follows
\begin{equation} 
  \label{eq:correlator} 
  K(\tau)=\frac{Z(\tau)}{Z(0)},
\end{equation} 
where $Z(\tau)$ is the functional integral given by
\begin{eqnarray} 
  \label{eq:funcint} 
  Z(\tau)&=& \int \exp[-{\cal  S}_{0} - {\cal  S}_C (\tau) -  
             {\cal S}' ]{\cal D}\phi(x,t).
\end{eqnarray}
Here ${\cal S}_{0}$ denotes the part of the Eucledian action derived from
the Hamiltonian (\ref{eq:freehamiltonian}) of free electrons moving
through the constriction in the absence of both interactions and
backscattering:
\begin{mathletters}
\label{eq:actionspinless}
\begin{equation}
  \label{eq:S_0}
  {\cal S}_{0}= \int_0^\beta dt \int dx \frac{v_F}{2\pi} \left[
           \frac{(\partial_t\phi)^2}{v_F^2}+(\partial_x \phi)^2\right].
\end{equation}
Here $\beta=1/T$.  The term $S_C(\tau)$ is the part of the action which is
due to the charging energy (\ref{eq:H_Cspinless}), where the operator
$\hat n$ is replaced with (\ref{eq:ntau}),
\begin{equation}
  \label{eq:S_C}
  {\cal S}_{C}(\tau)=\int_0^\beta  E_C \left[n_\tau (t) + 
                     \frac{1}{\pi}\phi(0,t) - N\right]^2 dt.
\end{equation}
Finally, $S'$ is the small contribution to the action due to the
backscattering (\ref{eq:scatteringspinless}),
\begin{equation}
  \label{eq:S'}
  S' = -\int_0^\beta \frac{D}{\pi}|r| \cos[2\phi(0,t)]\, dt.
\end{equation}
\end{mathletters}

The following calculations are performed in the regime of low
temperatures, $T\ll E_C$.  At $r=0$ the functional integral
(\ref{eq:funcint}) is gaussian, and its explicit evaluation gives
\begin{equation}
  \label{eq:K_0spinless}
  K_0 (\tau) = \left(\frac{\pi^2T}{\gamma E_C}\right)^2
               \frac{1}{\sin^2(\pi T\tau)},
\end{equation}
see Appendix~\ref{sec:ksl}. Here $\gamma=e^{\bf C}$, where ${\bf
  C}\approx0.577$ is Euler's constant.  The substitution of
(\ref{eq:K_0spinless}) into the expression (\ref{eq:thermocondk}) results
in $G_T=0$, as the integrand is an odd function of $t$.  This is a
consequence of the fact that at $r=0$ the system described by the
Hamiltonian (\ref{eq:freehamiltonian}) and (\ref{eq:H_Cspinless})
possesses electron-hole symmetry.  Indeed, in the absence of
backscattering the dependence of the Hamiltonian on the gate voltage $N$
can be removed by shifting the field $\phi(x)\to \phi(x)+\pi N$.  Then the
Hamiltonian $H_R^{(0)}+H_C$ is obviously invariant with respect to the
electron-hole symmetry transformation $\hat n \to -\hat n$, $\phi\to
-\phi$, and $\Pi\to-\Pi$.  On the other hand, the shift $\phi(x)\to
\phi(x)+\pi N$ will change the backscattering term
(\ref{eq:scatteringspinless}).  Thus the backscattering breaks the
electron-hole symmetry and gives rise to non-zero thermopower of the
device.

To account for the small backscattering at $|r|\ll1$ we expand the action
in Eqs.~(\ref{eq:correlator}) and (\ref{eq:funcint}) to first order in
$S'$ and find 
\begin{equation} 
\label{eq:kptsl} 
   K(\tau) = K_0(\tau) (1-\langle {\cal S}'\rangle_\tau  
             + \langle {\cal S}'\rangle_0).
\end{equation} 
Here $\langle {\cal S}'\rangle_\tau$ is defined as
\begin{equation} 
\label{eq:avdef} 
\langle {\cal S}'\rangle_\tau = 
   \frac{\int {\cal S}' \exp[-{\cal S}_{0}-{\cal S}_{C}(\tau)]{\cal D}\phi}
        {\int \exp[-{\cal S}_{0}-{\cal S}_{C}(\tau)]{\cal D}\phi}.
\end{equation} 
 
Due to the form (\ref{eq:S'}) of the perturbation ${\cal S}'$ the
evaluation of $K(\tau)$ in Eq.~(\ref{eq:kptsl}) again amounts to taking
gaussian functional integrals. The straightforward but lengthy
calculations carried out in Appendix~\ref{sec:ksl} give the result
\begin{eqnarray} 
 K(\tau) = K_{0}(\tau) 
           \big[&&1 - 2\gamma \xi |r| \cos (2\pi N)\nonumber\\
          &&+ 4\pi^2\xi\gamma|r|{\textstyle\frac{T}{E_C}}
                \sin(2\pi N)\cot(\pi T\tau)\big].
\label{eq:kslresult} 
\end{eqnarray} 
Here $\xi\approx1.59$ is a constant defined by Eq.~(\ref{eq:xi}).
The substitution of this result into Eq.~(\ref{eq:thermocondk}) gives the
following result for the thermoelectric coefficient in the first order in
$|r|$, 
\begin{equation} 
\label{eq:thermocondslresult} 
G_{T} = -\frac{8 \pi^7 \xi G_L}{15\gamma e}
         \left( \frac{T}{E_C}\right)^3 |r| \sin(2\pi N).
\end{equation} 

It is interesting to note that the thermoelectric coefficient $G_T$ given
by Eq.~(\ref{eq:thermocondslresult}) is an odd function of the gate
voltage $N$.  This property is more general than the perturbative result
(\ref{eq:thermocondslresult}).  Indeed, one can see from the form of the
functional integral (\ref{eq:funcint}) and the action
(\ref{eq:actionspinless}) that the correlator (\ref{eq:correlator}) has
the following symmetry property: $K(\beta-\tau,N)=K(\tau,1-N)$.
Furthermore, all the physical properties of the system are periodic in $N$
with period~1.  This can be shown by shifting $\phi\to\phi+\pi N$ which
removes $N$ from $S_C(\tau)$ and changes the cosine in Eq.~(\ref{eq:S'})
to $\cos[2\phi(0,t)+2\pi N]$.  The action then becomes invariant with
respect to the shift $N\to N+1$.  Consequently the correlator
(\ref{eq:correlator}) has the property $K(\beta-\tau,N)=K(\tau,-N)$.  One
can easily see from Eq.~(\ref{eq:thermocondk}) that only the part of
$K(\tau)$ which is odd with respect to $\tau\to\beta-\tau$ contributes to
the thermoelectric coefficient $G_T$.  Using the aforementioned properties
of $K(\tau,N)$, this odd part can be presented as
\begin{eqnarray}
  \label{eq:K_odd}
  K_{\rm odd}(\tau)&=&\frac12[K(\tau,N)-K(\beta-\tau,N)]
  \nonumber\\
                   &=&\frac12[K(\tau,N)-K(\tau,-N)].
\end{eqnarray}
Therefore the thermoelectric coefficient $G_T$ is an odd function of $N$.

The conductance $G$ can be obtained by substitution of correlator
(\ref{eq:kslresult}) analytically continued to real time into Eq.~(52) of
Ref.~\onlinecite{Furusaki95}.  The result has the form
\begin{equation}
  \label{eq:G_asymmetric}
  G = G_L\frac{2\pi^4 T^2}{3\gamma^2 E_C^2}
         \big[1 - 2\gamma \xi |r| \cos (2\pi N)\big].
\end{equation}
This expression is in agreement with the formula (A27) of
Ref.~\onlinecite{Furusaki95}, where the numerical prefactor in the
brackets was not determined, and with the expression (34) of
Ref.~\onlinecite{Yi96}, where the constant $\xi$ was found.

Substituting Eq.~(\ref{eq:thermocondslresult}) into
Eq.~(\ref{eq:thermopowerdef}) and using the leading term in
Eq.~(\ref{eq:G_asymmetric}) for the conductance, we obtain the following
expression for the thermopower in the spinless case
\begin{equation} 
\label{eq:S_spinless} 
S = -\frac{4\pi^3\xi\gamma T}{5 e E_C}|r|\sin(2\pi N). 
\end{equation} 

It is instructive to compare this result with the thermopower
(\ref{eq:S_weak}) in the regime of weak inelastic cotunneling.  Both
expressions vanish linearly at $T\to0$, but unlike Eq.~(\ref{eq:S_weak}),
our result (\ref{eq:S_spinless}) depends on the transmission
coefficient of the barrier.  As expected, at perfect transmission $r\to0$
the Coulomb blockade oscillations of the thermopower disappear.

\subsection{Electrons with spin} 
\label{sec:spinful}

Although the spins of electrons can be polarized in an experiment by
applying a strong magnetic field, the most common situation is when no
field is applied.  In this regime one has to consider the case of
spin-$\frac12$ electrons.

In the presence of electron spins the Hamiltonian
(\ref{eq:freehamiltonian})--(\ref{eq:H_Cspinless}) has to be modified to
account for the two species of electrons: spin-$\uparrow$ and
spin-$\downarrow$.  Each of the spin subsystems can be bosonized
independently, and the Hamiltonian takes the form\cite{Matveev95}
\begin{mathletters}
\begin{eqnarray}
  \label{eq:freehamiltonian_spins}
    H_R^{(0)}&=&\frac{v_F}{2\pi} \sum_{\sigma=\uparrow,\downarrow}
             \int_{-\infty}^\infty 
             \left\{\pi^2\Pi_\sigma^2(x)+
              [\partial_x\phi_\sigma(x)]^2\right\}dx ,\\ 
  \label{eq:scatteringspins}
  H_R' &=&  -\frac{D}{\pi}|r|\left\{\cos\left[2\phi_\uparrow(0)\right]
                       +\cos\left[2\phi_\downarrow(0)\right]\right\},\\
  \label{eq:H_Cspins}
   H_C &=& E_C \left\{\hat{n} + 
          \frac{1}{\pi}\big[\phi_\uparrow(0)
            +\phi_\downarrow(0)\big] - N \right\}^2.
\end{eqnarray}
\end{mathletters}
To find the thermoelectric coefficient (\ref{eq:thermocondk}) one has to
find the correlator $K(\tau)$.  Similarly to the case of spinless
electrons, $K(\tau)$ can be expressed in terms of the imaginary-time
functional integral (\ref{eq:correlator}), where 
\begin{eqnarray} 
  \label{eq:funcint_spins} 
  Z(\tau)&=& \int \exp[-{\cal  S}_{0} - {\cal  S}_C (\tau) -  
             {\cal S}' ]{\cal D}\phi_c {\cal D}\phi_s.
\end{eqnarray}  
Here we have introduced the charge and spin fields
$\phi_{c,s}(x,t)=[\phi_\uparrow(x,t)\pm\phi_\downarrow(x,t)]/\sqrt2$.  The
action in Eq.~(\ref{eq:funcint_spins}) is expressed in terms of these
variables as
\begin{mathletters} 
  \label{eq:actionspin}
\begin{eqnarray} 
  \label{eq:actionspina} 
&&{\cal S}_{0}= \sum_{\alpha=c,s}\int_0^\beta dt \int dx 
                \frac{v_F}{2\pi}
                \left[\frac{(\partial_t\phi_\alpha)^2}{v_F^2}
                     +(\partial_x \phi_\alpha)^2\right], \\ 
 \label{eq:actionspinb} 
&&{\cal S}_{C}(\tau)=\int_0^\beta dt E_C \left[ n_\tau (t) + 
  \frac{\sqrt{2}}{\pi}\phi_c  
  (0,t) -N\right]^2, \\ 
 \label{eq:actionspinc} 
&&{\cal S}' =-\int_0^\beta dt \frac{2D}{\pi}|r|\cos[\sqrt{2}\phi_c(0,t)] 
\cos[\sqrt{2}\phi_s(0,t)]. 
\end{eqnarray} 
\end{mathletters} 

Similar to the case of spinless electrons, in the absence of
backscattering in the constriction the calculation of $K(\tau)$ reduces to
evaluation of a gaussian functional integral (\ref{eq:funcint_spins}).
Clearly, at $r=0$ the integral over $\phi_s$ is unaffected by $n_\tau(t)$;
therefore the integrals over the spin degrees of freedom in the numerator
and denominator of Eq.~(\ref{eq:correlator}) cancel.  One can easily see
that the action (\ref{eq:actionspina}) and (\ref{eq:actionspinb}) of the
charge mode is identical to that of the spinless problem (\ref{eq:S_0})
and (\ref{eq:S_C}) upon the substitution $E_C\to 2E_C$, $n_\tau(t)\to
n_\tau(t)/\sqrt2$, and $N\to N/\sqrt2$.  Making the respective
modifications to the derivation of $K_0(\tau)$ in Appendix~\ref{sec:ksl1},
we find
\begin{equation}
  \label{eq:K_0spins}
  K_c (\tau) = \frac{\pi^2T}{2\gamma E_C}
               \frac{1}{|\sin(\pi T\tau)|}.
\end{equation}
Substituting the analytic continuation of this result to $\tau=1/2T+it$
into Eq.~(\ref{eq:thermocondk}), we find $G_T=0$.  As it was explained in
Sec.~\ref{sec:spinless}, this is the consequence of the fact that the
system possesses electron-hole symmetry at $r=0$.

The rest of this section is organized as follows.  In
Sec.~\ref{sec:spinfulpert} we calculate the thermopower within the
second-order perturbation theory in the reflection amplitude $r$.  We show
that the perturbative result diverges at low temperatures.  We then find
the thermopower at arbitrarily low temperatures in
Sec.~\ref{sec:nonperturbative} using a non-perturbative approach.

\subsubsection{Perturbation theory}
\label{sec:spinfulpert}

At non-vanishing backscattering the correction to $K_c(\tau)$ appears in
the second order in $r$.  Indeed, the first-order correction can be
expressed in the form (\ref{eq:kptsl}).  It is easy to check that unlike
the case of spinless electrons, the average $\langle{\cal S}'\rangle$
vanishes, because the fluctuations of the spin mode $\phi_s(0,t)$ are not
suppressed at low frequencies by the charging energy term
(\ref{eq:actionspinb}).  Expanding Eqs.~(\ref{eq:correlator}) and
(\ref{eq:funcint_spins}) to second order in $r$, we find
\begin{equation} 
\label{eq:kptsp} 
   K(\tau) = K_c(\tau) \left[1
             + \frac12\left(\langle {\cal S}'^2\rangle_\tau  
             - \langle {\cal S}'^2\rangle_0\right)\right].
\end{equation} 
Similarly to Eq.~(\ref{eq:avdef}), the averaging
$\langle\ldots\rangle_\tau$ here is performed with the action ${\cal
  S}_0+{\cal S}_C(\tau)$ given by (\ref{eq:actionspina}) and
(\ref{eq:actionspinb}).  Using the explicit form (\ref{eq:actionspinc}) of
${\cal S}'$, we get
\begin{equation}
  \label{eq:S_squared}
  \langle {\cal S}'^2\rangle_\tau = \frac{4D^2|r|^2}{\pi^2}
            \int_0^\beta\int_0^\beta 
            \kappa_c(t,t';\tau)\kappa_s(t,t')dt\,dt',
\end{equation}
where we have introduced the correlators
\begin{mathletters}
\label{eq:kappas}
\begin{eqnarray}
  \label{eq:kappa_c}
  \kappa_c(t,t';\tau)&=&\langle \cos[\sqrt{2}\phi_c(0,t)]
                                \cos[\sqrt{2}\phi_c(0,t')] 
                        \rangle_\tau,\\
  \label{eq:kappa_s}
  \kappa_s(t,t')&=&\langle \cos[\sqrt{2}\phi_s(0,t)]
                                \cos[\sqrt{2}\phi_s(0,t')] \rangle_0.
\end{eqnarray}
\end{mathletters}
The spin fluctuations are completely decoupled from the charging action
(\ref{eq:actionspinb}), rendering the correlator $\kappa_s$ independent of
$\tau$.  The calculation of the correlators $\kappa_c$ and $\kappa_s$
reduces to evaluation of gaussian integrals.  In Appendix~\ref{sec:kappas}
we find
\begin{eqnarray}
  \label{eq:kappa_c_result}
  \kappa_c(t,t';\tau)&=&\frac{\gamma E_C}{\pi D}{\rm Re}
              \left(e^{2i\pi N}e^{-i[\chi_\tau(t)+\chi_\tau(t')]}\right.
              \nonumber\\
           &&\qquad\qquad\left.+e^{-i[\chi_\tau(t)-\chi_\tau(t')]}\right),\\
  \kappa_s(t,t')&=&\frac{\pi T}{2D}\frac{1}{|\sin[\pi T(t-t')]|},
  \label{eq:kappa_s_result}
\end{eqnarray}
where we have introduced the notation
\begin{mathletters}
  \label{eq:chi}
\begin{eqnarray}
  \label{eq:chi_tau_expansion}
  \chi_\tau(t)&=&\pi n_\tau(t) + \delta\chi_\tau(t),
\\
  \label{eq:delta_chi}
  \delta\chi_\tau(t)&=&\sum_{n=1}^\infty
                     \frac{\sin[2\pi T(t-\tau)n]-\sin[2\pi Ttn]}
                          {n+\frac{E_C}{\pi^2T}}.
\end{eqnarray}
\end{mathletters}
In Eq.~(\ref{eq:kappa_c_result}) we have assumed $|t-t'|\gg E_C^{-1}$; we
will see that this region gives the leading contribution to the integral
(\ref{eq:S_squared}). 

As discussed in Sec.~\ref{sec:spinless}, only the odd in $N$ part
(\ref{eq:K_odd}) of the correlator $K(\tau)$ contributes to the
thermoelectric coefficient (\ref{eq:thermocondk}).  Keeping only the odd
part of Eq.~(\ref{eq:kappa_c_result}), from (\ref{eq:kptsp}) in the second
order in $r$ we find
\begin{eqnarray}
  \label{eq:K_odd_integral}
  K_{\rm odd}(\tau)&=&K_c(\tau)\frac{2\gamma E_C T}{\pi^2}
                      |r|^2\sin(2\pi N){\cal I}(\tau),\\
     {\cal I}(\tau)&=&\int_0^\beta dt \sin\chi_\tau(t)
                   \int_0^\beta \frac{\cos\chi_\tau(t')}
                                     {|\sin[\pi T(t-t')]|}dt'.
  \label{eq:cal_I}
\end{eqnarray}
In evaluating the integral ${\cal I}(\tau)$ one should keep in mind that
the denominator in Eq.~(\ref{eq:cal_I}) is written for $|t-t'|\gg
E_C^{-1}$.  Thus the logarithmic divergence at $t=t'$ should be cut off at
$|t-t'|\sim E_C^{-1}$. 

To evaluate the integral ${\cal I}(\tau)$ we first notice that away from
the points $t=0,\tau,\beta$ the correction $\delta\chi_\tau(t)$ in
Eq.~(\ref{eq:chi}) is small in $T/E_C$,
\begin{equation}
  \label{eq:delta_chi_approximate}
  \delta\chi_\tau(t)\simeq\frac{\pi^2T}{2E_C}
                     \{\cot[\pi T(t-\tau)]-\cot[\pi Tt]\}.
\end{equation}
One can neglect this correction in the argument of the cosine in
Eq.~(\ref{eq:cal_I}) and replace $\cos\chi_\tau(t')={\rm sgn}(t'-\tau)$.
Then the integral over $t'$ can be evaluated with logarithmic accuracy:
\begin{eqnarray*}
  {\cal I}(\tau)&\simeq&\frac{2}{\pi T}\int_0^\beta dt \sin\chi_\tau(t)
                   \left[
                    -\ln\tan\frac{\pi Tt}{2}\right.\\
                 &&\qquad\qquad   \left.
                  +\,{\rm sgn}(t-\tau)
                      \ln\left(\frac{E_C}{T}
                               \tan\frac{\pi T|t-\tau|}{2}\right)
                   \right].
\end{eqnarray*}
To leading order in $T/E_C$ one can replace
$\sin\chi_\tau(t)=\delta\chi_\tau(t) \,{\rm sgn}(t-\tau)$.  Using the
approximation (\ref{eq:delta_chi_approximate}) we then obtain with
logarithmic accuracy
\begin{equation}
  \label{eq:K_odd_result}
  K_{\rm odd}(\tau)=-K_c(\tau)\frac{8\gamma}{\pi^2}
                      |r|^2\sin(2\pi N)\ln\frac{E_C}{T}
                      \ln\tan\frac{\pi T\tau}{2}.
\end{equation}
Substituting this result for $K(\tau)$ in Eq.~(\ref{eq:thermocondk}) we
find 
\begin{equation}
  \label{eq:G_T_spins_perturbation}
  G_T=-\frac{8\pi}{9}\frac{G_L}{e}\frac{T}{E_C}\ln\frac{E_C}{T}\,
                               |r|^2\sin(2\pi N).
\end{equation}
The second-order perturbation theory result for the thermopower can be
found from Eq.~(\ref{eq:thermopowerdef}) using the result
$G=G_L\frac{\pi^3T}{8\gamma E_C}$ of Ref.~\onlinecite{Furusaki95} for the
conductance of the device at $r=0$,
\begin{equation}
  \label{eq:S_spins_perturbative_result}
  S=-\frac{64\gamma}{9\pi^2}\frac{1}{e}\ln\frac{E_C}{T}\,
                               |r|^2\sin(2\pi N).
\end{equation}
This result applies at $T\ll E_C$ and, similarly to the spinless case
(\ref{eq:S_spinless}), the thermopower vanishes at $r\to 0$.  It is
important to note that unlike the spinless case (\ref{eq:S_spinless}), the
thermopower (\ref{eq:S_spins_perturbative_result}) diverges at $T\to0$.
This means, in particular, that the perturbation theory leading to
Eq.~(\ref{eq:S_spins_perturbative_result}) fails at sufficiently low
temperatures.  In the next section we perform a non-perturbative
calculation and establish the true behavior of the thermopower at $T\to0$.

\subsubsection{Non-perturbative treatment}
\label{sec:nonperturbative}

The logarithmic growth of the thermopower
(\ref{eq:S_spins_perturbative_result}) at low temperature indicates that
the thermoelectric properties of the system are controlled by the
spin and charge fluctuations at frequencies below $E_C$.  In this section
we construct a theory that describes the  low-energy properties of the
system exactly and enables us to obtain a non-perturbative expression for
the thermopower at arbitrarily low temperatures.  This derivation was
outlined in Ref.~\onlinecite{PRL01}. 

As we already discussed, at $r=0$ the contributions of the spin
fluctuations to the functional integrals in the numerator and the
denominator of Eq.~(\ref{eq:correlator}) cancel each other, and the ratio
of the functional integrals over the charge degrees of freedom is equal to
the correlator $K_c(\tau)$, Eq.~(\ref{eq:K_0spins}).  The effect of small
but finite $r$ on the charge modes is negligible, because their
fluctuations at low energies are suppressed by the charging energy.
However, even a small backscattering $r$ pins the fluctuations of the spin
modes and changes their low-frequency dynamics
dramatically.\cite{Furusaki95} Therefore one can account for the small
backscattering by presenting the correlator (\ref{eq:correlator}) in the
form
\begin{equation}
  \label{eq:correlator_effective}
  K(\tau)=K_c(\tau)K_s(\tau),\quad K_s(\tau)=\frac{Z_s(\tau)}{Z_s(0)},
\end{equation}
where $Z_s(\tau)$ is the functional integral over the slow spin
modes, averaged over the fast charge modes.

The calculation of $Z_s(\tau)$ amounts to integrating out the fast charge
degrees of freedom in the functional integral (\ref{eq:funcint_spins}).
Since the spin and charge fluctuations are only coupled by the
backscattering term (\ref{eq:actionspinc}), this procedure reduces to the
averaging of $\cos[\sqrt{2}\phi_c(0,t)]$ with the gaussian action ${\cal
  S}_0+{\cal S}_C(\tau)$.  Indeed, one can rewrite Eqs.~(\ref{eq:kappa_c})
and (\ref{eq:kappa_c_result}) as
\begin{eqnarray}
  \label{eq:decoupling}
   &\langle \cos[\sqrt{2}\phi_c(0,t)]
          \cos[\sqrt{2}\phi_c(0,t')]\rangle_\tau =
       \Upsilon_\tau(t)\Upsilon_\tau(t'),&\\[1ex]
   &\Upsilon_\tau(t)=\sqrt{\frac{2\gamma E_C}{\pi D}} 
                    \cos[\chi_\tau(t)-\pi N].&
\label{eq:Upsilon}
\end{eqnarray}
One can see from Eq.~(\ref{eq:decoupling}) that at $|t-t'|\gg E_C^{-1}$
the correlator (\ref{eq:kappa_c}) factorizes into the product of the
averages of the cosines, and that $\langle
\cos[\sqrt{2}\phi_c(0,t)]\rangle_\tau=\Upsilon_\tau(t)$.  It is clear that
the higher-point correlators will also factorize into the product of
averages.  One can therefore simply replace the
$\cos[\sqrt{2}\phi_c(0,t)]$ in the action (\ref{eq:actionspinc}) by
$\Upsilon_\tau(t)$ and obtain the effective action for the spin degrees of
freedom in the form
\begin{eqnarray}
  {\cal S}_\tau &=&  \int_0^\beta dt \int dx 
                \frac{v_F}{2\pi}
                \left[\frac{(\partial_t\phi_s)^2}{v_F^2}
                     +(\partial_x \phi_s)^2\right] \nonumber\\
                &&-\int_0^\beta \sqrt\frac{4D}{v_F}\,\lambda_\tau(t)
                   \cos[\sqrt{2}\phi_s(0,t)]dt,
  \label{eq:effective_action}
\end{eqnarray}
where we have introduced the notation
\begin{eqnarray}
   \lambda_\tau(t)&=&\Lambda
                     \cos[\chi_\tau(t)-\pi N]\nonumber\\
\label{eq:lambda_tau}
                 &=&\Lambda (-1)^{n_\tau(t)}
                     \cos[\delta\chi_\tau(t)-\pi N],\\
   \Lambda&=&\sqrt{\case{2\gamma v_F E_C}{\pi^3}}\, |r|.
\label{eq:Lambda}
\end{eqnarray}
The procedure leading to the action (\ref{eq:effective_action}) implied
that all the relevant time scales of the problem are longer than
$E_C^{-1}$.  Therefore one has to integrate out the fluctuations of the
spin degrees of freedom with frequencies exceeding $E_C$.  This procedure
is straightforward and amounts to replacing $D$ with the new bandwidth
$\sim E_C$.  Thus Eq.~(\ref{eq:effective_action}) gives the effective
action of the problem, provided the bandwidth $D\sim E_C$.

One can now find the correlator $K(\tau)$ using
Eq.~(\ref{eq:correlator_effective}) where the functional integral
$Z_s(\tau)$ is defined as $Z_s(\tau)=\int e^{-{\cal S}_\tau}{\cal
  D}\phi_s$.  For the subsequent calculations it will be convenient to use
the hamiltonian formulation of the problem and express $Z_s(\tau)$ as the
trace of the time-ordered exponential:
\begin{equation}
  \label{eq:trace}
  Z_s(\tau)= {\rm Tr} \left\{T_t \exp\left[
                    -\int_0^\beta H_\tau(t)dt\right]\right\},
\end{equation}
where the time dependent Hamiltonian $H_\tau$ is given by
\begin{eqnarray}
  H_\tau(t)&=&\frac{v_F}{2\pi} \int_{-\infty}^\infty 
             \left\{\pi^2\Pi_s^2(x)+
              [\partial_x\phi_s(x)]^2\right\}dx \nonumber\\
           && - \sqrt\frac{4D}{v_F}\,\lambda_\tau(t)
                \cos[\sqrt{2}\phi_s(0)].
  \label{eq:H_tau}
\end{eqnarray}

The small parameter of the problem $r$ enters through $\lambda_\tau(t)$.
In order to evaluate $K(\tau)$ in all orders in $\lambda_\tau$ we
refermionize the Hamiltonian (\ref{eq:H_tau}) following
Ref.~\onlinecite{Matveev95} and find
\begin{equation}
  \label{eq:H_tau_fermionized}
  H_\tau(t)=
  \int_{-\infty}^\infty \left[\xi_k c_k^\dagger c_k
         -\lambda_\tau(t)(c+c^\dagger)(c_k-c_k^\dagger)\right]dk.
\end{equation}
Here $\xi_k=v_F k$; the operators $c_k^\dagger$ and $c_k$ satisfying the
anticommutation relations $\{c_k,c_{k'}^\dagger\}=\delta(k-k')$ create and
destroy chiral fermions.  Finally, $c$ is a fermion annihilation operator
anticommuting with $c_k$ and $c_k^\dagger$.

Although the Hamiltonian (\ref{eq:H_tau_fermionized}) is quadratic in the
fermion operators, the time dependence of $\lambda_\tau(t)$ makes the
evaluation of the trace (\ref{eq:trace}) non-trivial.  It is clear from
Eq.~(\ref{eq:delta_chi}) that at $T/E_C\ll 1$ the main time dependence is
due to the factor $(-1)^{n_\tau(t)}$ in the definition of
$\lambda_\tau(t)$, Eq.~(\ref{eq:lambda_tau}).  One can greatly simplify
the calculation by eliminating this time dependence with the following
trick. Note that the unitary transformation with the operator
\begin{equation}
  \label{eq:unitary}
  U=(-1)^{c^\dagger c}=(c-c^\dagger)(c+c^\dagger)
\end{equation}
changes the sign of $\lambda_\tau$ in the Hamiltonian
(\ref{eq:H_tau_fermionized}).  Therefore the factor $(-1)^{n_\tau(t)}$ can
be accounted for by adding operators $U(\tau)$ and $U(0)$ to the trace
(\ref{eq:trace}), 
\begin{equation}
  \label{eq:trace_U}
    Z_s(\tau)= {\rm Tr} \left\{T_t \exp\left[
                    -\int_0^\beta [H_0+H'_\tau(t)]dt\right]
                     U(\tau)U(0)\right\}.
\end{equation}
Here $H_0+H'_\tau(t)$ is obtained from the Hamiltonian
(\ref{eq:H_tau_fermionized}) by replacing
$\lambda_\tau(t)\to\lambda_\tau(t)/(-1)^{n_\tau(t)}$.  Its
time-independent part $H_0$ is given by (\ref{eq:H_tau_fermionized}) at
$\tau=0$, and the correction is
\begin{eqnarray}
  \label{eq:H'_tau}
  H'_\tau(t) &=& \Lambda\{\cos(\pi N)
                        -\cos[\delta\chi_\tau(t)-\pi N]\}
       (c+c^\dagger)\Psi,\\[1ex]
  \Psi &=& \int_{-\infty}^\infty (c_k-c_k^\dagger)dk.
  \label{eq:Psi}
\end{eqnarray}

The perturbation $H'_\tau(t)$ vanishes at $T/E_C\to0$.  In this limit the
spin contribution to the correlator (\ref{eq:correlator_effective})
becomes the Green's function of operators $U$,
\begin{equation}
  \label{eq:GreenU}
  K_s^{(0)}(\tau)=\langle T_t U(\tau)U(0)\rangle_0,
\end{equation}
where $\langle\ldots\rangle_0$ denotes averaging over the equilibrium
thermal distribution with the Hamiltonian $H_0$.  The explicit analytic
result for this quantity is given by formula (66) of
Ref.~\onlinecite{Furusaki95}.  The result is an even function of the gate
voltage $N$, and therefore within the approximation $H'_\tau(t)=0$ the
thermoelectric coefficient $G_T$ vanishes.

To find the leading contribution to $G_T$ at small $T/E_C$ we expand
Eq.~(\ref{eq:trace_U}) to first order of the perturbation theory in 
$H'_\tau(t)$.  The correction to $K_s(\tau)$ has the form
\begin{equation}
  \label{eq:K_s_first_order}
  K_s^{(1)}(\tau)=-\int_0^\beta dt\,
                  \langle T_t H'_\tau(t) U(\tau)U(0)\rangle_0.
\end{equation}
This correction is evaluated with logarithmic accuracy in
Appendix~\ref{sec:K_s_nonperturbative},
\begin{eqnarray}
  \label{eq:K_s_nonperturbative_result}
  K_s^{(1)}(\tau) &=& -\frac{8\gamma}{\pi^{2}}
                |r|^2\sin(2\pi N)\ln\frac{E_C}{T+\Gamma}\nonumber \\ 
   &&\times  \int_{-\infty}^\infty
      \frac{\xi d\xi}{\xi^2+\Gamma^2}
                \frac{e^{\xi\tau}}{e^{\beta \xi}+1},
    \\[1ex]
  \Gamma &=& \frac{8\gamma E_C}{\pi^2}|r|^2 \cos^2(\pi N).
  \label{eq:Gamma}
\end{eqnarray}
It is important to note that although this result is the first-order
correction in $H'_\tau(t)$, it is non-perturbative in the reflection
amplitude $r$.

Substituting the correlator $K(\tau)$ in the form
(\ref{eq:correlator_effective}) with $K_c$ and $K_s$ given by
Eqs.~(\ref{eq:K_0spins}) and (\ref{eq:K_s_nonperturbative_result}) into
the expression for the thermoelectric coefficient (\ref{eq:thermocondk}),
we obtain
\begin{eqnarray} 
G_T &=& -\frac{1}{6\pi}\frac{G_L}{e } \frac{T}{E_C}
         \ln{\frac{E_C}{T+\Gamma}}\,  |r|^2 \sin(2\pi N) 
\nonumber \\ 
&&\times 
\int_{-\infty}^{\infty}\frac{x^2(x^2+\pi^2)dx}
    {[x^2+(\Gamma/T)^2]\cosh ^2(x/2)}. 
\label{eq:result}
\end{eqnarray} 
At temperatures $T\gg\Gamma$ this expression reproduces the perturbative
result (\ref{eq:G_T_spins_perturbation}).  The latter is valid until
$T\sim \Gamma$, and at $T\ll\Gamma$ the thermoelectric coefficient $G_T$
becomes
\begin{equation}
  \label{eq:G_T_low_T}
  G_T= -\frac{\pi^7}{60\gamma^2}\frac{G_L}{e } \frac{T^3}{E_C^3}
          \frac{1}{|r|^2} 
          \frac{\sin(\pi N)}{\cos^3(\pi N)}
          \ln\frac{1}{|r|^2\cos^2(\pi N)}.
\end{equation}
The dependence of $G_T$ on the small reflection amplitude illustrates the
non-perturbative nature of this result.  It is also worth noting that at
low temperatures the dependence of $G_T$ on the gate voltage $N$ is
strongly non-sinusoidal.

To find the thermopower $S=G_T/G$ one can use the expression
(\ref{eq:result}) and the non-perturbative result of
Ref.~\onlinecite{Furusaki95} for the conductance $G$ of the SET,
\begin{equation} 
  \label{eq:conductance} 
  G=\frac{ G_L \Gamma}{8 \gamma 
    E_C}\int_{-\infty}^{\infty} 
  \frac{(x^2+\pi^2)dx}{\left[x^2+(\Gamma/T)^2\right]\cosh ^2(x/2)}.
\end{equation} 
At relatively high temperatures $T\gg \Gamma$ the thermopower $S=G_T/G$
obtained from Eqs.~(\ref{eq:result}) and (\ref{eq:conductance}) coincides
with the perturbative expression (\ref{eq:S_spins_perturbative_result}).
In the more interesting case of low temperatures $T\ll \Gamma$, we find
\begin{equation} 
  \label{eq:limit} 
  S= -\frac{\pi^3}{5}\frac{1}{e}\frac{T}{E_C}\tan (\pi N)
      \ln\frac{1}{|r|^2\cos^2(\pi N)}.
\end{equation} 

The new energy scale $\Gamma$ arising from the non-perturbative solution
is always small compared to the charging energy, see Eq.~(\ref{eq:Gamma}).
It is important to keep in mind that $\Gamma$ is a function of the gate
voltage, and vanishes near the Coulomb blockade peaks,
$N=\pm\frac12,\pm\frac32,\pm\frac52,\ldots$.  As a result, even at $T\ll
E_C|r|^2$ the perturbative results (\ref{eq:G_T_spins_perturbation}) and
(\ref{eq:S_spins_perturbative_result}) are still valid near the
conductance peaks, whereas in the valleys the new asymptotics
(\ref{eq:G_T_low_T}) and (\ref{eq:limit}) apply.  The crossover between
these asymptotics occurs at the values of $N$ where $\Gamma=T$, i.e.,
according to Eq.~(\ref{eq:Gamma}) at a distance $\delta
N\sim\sqrt{T/E_C|r|^2}$ from the centers of the conductance peaks.  At
these points the thermopower reaches its maximum absolute value $S_{\rm
  max}$, which can be estimated by substituting $N=\frac12+\delta N$ in
either Eq.~(\ref{eq:S_spins_perturbative_result}) or Eq.~(\ref{eq:limit}),
resulting in
\begin{equation}
  \label{eq:S_max}
  S_{\rm max}\sim e^{-1} |r| \sqrt{\frac{T}{E_C}}\, \ln\frac{E_C}{T}.
\end{equation}

\begin{figure}[tbp]
  \begin{center}
   \epsfbox{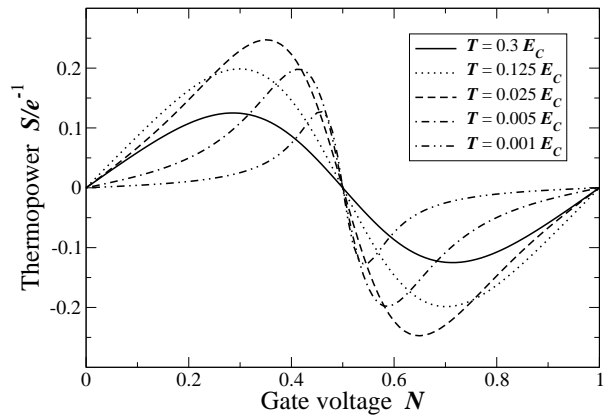}\\[2ex]
    \caption{{Thermopower of SET as a function of gate voltage at
        different temperatures. The curves for $S=G_T/G$ are obtained
        numerically from Eqs.~(\ref{eq:result}) and (\ref{eq:conductance})
        at $|r|^2=0.1$ and $T/E_C=0.3$, 0.125, 0.025, 0.005, 0.001. As the
        temperature is lowered, the amplitude of the thermopower
        oscillations first grows according to
        Eq.~(\ref{eq:S_spins_perturbative_result}) and then decreases in
        agreement with Eq.~(\ref{eq:S_max}). Below the crossover
        temperature $\sim E_C |r|^2$ the shape of the oscillations becomes
        non-sinusoidal. }}
    \label{fig:illustration}
  \end{center}
\end{figure}

The exact shape of the Coulomb blockade oscillations of the thermopower
found from Eqs.~(\ref{eq:result}) and (\ref{eq:conductance}) is
illustrated in Fig.~\ref{fig:illustration}.

\section{Summary and discussion} 
\label{sec:discussion}

We presented a theory of the thermopower of single electron transistors in
the regime when the coupling of the quantum dot to one of the leads is
strong.  The theory is applicable to devices with relatively large dots,
where the effects of finite quantum level spacing can be neglected, and
the main transport mechanism is inelastic cotunneling.  Using the fact
that the coupling to one of the two leads is weak, we obtained the
expression (\ref{eq:thermocondk}) for the thermoelectric coefficient $G_T$
in terms of the correlator $K(\tau)$ describing the charging of the dot
strongly coupled to the other lead.  The general expression
(\ref{eq:thermocondk}) is applicable to contacts with arbitrary coupling.
We applied it to the case of coupling via a quantum point contact with a
single transverse mode and almost perfect transmission, $|r|\ll1$.  In the
case of spin-polarized electrons we found sinusoidal Coulomb blockade
oscillations of the thermopower with the amplitude $\sim e^{-1}|r| T/E_C$,
Eq.~(\ref{eq:S_spinless}).  Experimentally the polarization of electron
spins can be achieved by applying a strong magnetic field.  In the absence
of the magnetic field the thermopower is given by the ratio of
non-perturbative expressions (\ref{eq:result}) and (\ref{eq:conductance}).
At relatively high temperatures $T\gg E_C|r|^2$ the Coulomb blockade
oscillations of $S$ are sinusoidal, with the amplitude $\sim e^{-1}
|r|^2\ln(E_C/T)$, Eq.~(\ref{eq:S_spins_perturbative_result}).  At lower
temperatures $T\ll E_C|r|^2$ the oscillations are non-sinusoidal,
Fig.~\ref{fig:illustration}, and their amplitude is given by
Eq.~(\ref{eq:S_max}).

We are aware of only one experiment on the thermopower of SET in the
strong coupling regime, Ref.~\onlinecite{Molenkamp98}.  In this experiment
the Coulomb blockade oscillations of the thermopower $S(N)$ were measured
at different values of the reflection coefficient.  Only one published
curve $S(N)$, measured at $|r|^2= 0.2\pm0.1$ approached the strong
tunneling limit $|r|\ll 1$.  In this case the thermopower remained
sinusoidal even at the lowest available temperatures.  To observe the more
interesting non-sinusoidal behavior of $S(N)$ one would have to measure
the thermopower at lower temperature to reach the regime $T\ll E_C |r|^2$.
This may require making a sample with a larger quantum dot to ensure that
the lowest temperature is still large compared to the quantum level
spacing.

\acknowledgements

This research was supported by the NSF Grants No.~DMR-9974435 and
DMR-9984002, and by the Sloan and the Packard foundations.  We are also
grateful to B.L. Altshuler and L.~Molenkamp for valuable discussions.

\appendix 
 
\section{Tunneling density of states} 
\label{sec:dos} 
 
In this appendix we present the derivation of Eq.~(\ref{eq:dos}) for the
tunneling density of states $\nu (\epsilon)$.  We start with the standard
expression for the density of states
\begin{equation} 
\label{eq:dosret} 
\nu(\epsilon)=\frac{i}{2\pi}\left[G^R(\epsilon)-G^A(\epsilon)\right],
\end{equation} 
where $G^R$ and $G^A$ are the retarded and advanced Green's functions,
which can be obtained by the analytic continuation of the Matsubara
Green's function $G(\epsilon_n)$.

In the frequency representation the Matsubara Green's function can be
written as
\begin{equation} 
\label{eq:matsubara} 
G(\epsilon_n)=\int_0^\beta d\tau \exp(i\epsilon_n \tau)G(\tau),
\end{equation} 
where $\epsilon_n=\pi T(2n+1)$ are the fermionic Matsubara frequencies.
Depending on the sign of $\epsilon_n$ the $\tau$ integration contour can
be distorted to the upper or lower half plane as shown in
Fig.~\ref{fig:contour}.

\narrowtext{ 
\begin{figure} 
\begin{center} 
\epsfxsize=6cm 
\epsfbox{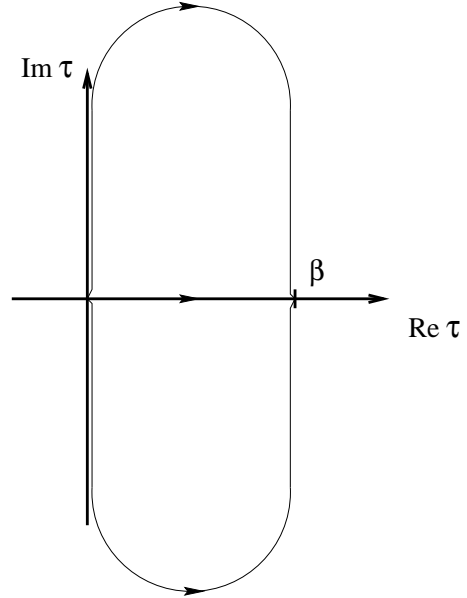} 
\end{center} 
\caption{ Representation of the deformation of the $\tau$-integration contour
  in Eq.~(\ref{eq:matsubara}). For positive $\epsilon_n$ the contour
  should be distorted into the upper half plane, and for negative
  $\epsilon_n$ --- into the lower half plane.  }
\label{fig:contour} 
\end{figure} 
}

Keeping in mind that the retarded Green's function $G^R(i\epsilon_n)=
G(\epsilon_n)$ at $\epsilon_n>0$, and using the fact that for the
fermionic Matsubara frequencies $\exp(i\epsilon_n \beta)=-1$, we can then
express the retarded Green function as
\begin{eqnarray} 
\label{eq:matsubaradist} 
G^{R}( i \epsilon_n)&=&i \int_0^{\infty}dt \exp (-\epsilon_n t) 
\nonumber \\ 
&& \qquad \times [ G( it +0)+G(\beta-0 
+it )]  .
\end{eqnarray} 
The analytic continuation to real frequencies can now be performed in the
last line of Eq.~(\ref{eq:matsubaradist}) through the substitution
$\epsilon_n\to -i \epsilon$.  

We then obtain a similar expression for the advanced Green's function
$G^A(\epsilon)$ using the relation $G^A(i\epsilon_n)= G(\epsilon_n)$ at
$\epsilon_n<0$.  Combining the two results, we find the following
expression for the density of states (\ref{eq:dosret})
\begin{eqnarray} 
\label{eq:dosdist} 
\nu(\epsilon )&=&-\frac{1}{2\pi} \int_{-\infty}^{\infty}dt  
\exp (i \epsilon t) G( it +0) \nonumber \\ 
&&-\frac{1}{2\pi} \int_{-\infty}^{\infty}dt  
\exp (i \epsilon t )G(\beta-0 
+it ). 
\end{eqnarray} 
Since the Green's function $G(\tau)$ is analytic everywhere except on the
lines ${\rm Re}\, \tau = 0,\pm\beta,\pm2\beta,\ldots$, we can shift the
integration contour in the first line of Eq.~(\ref{eq:dosdist}) by $t \to
t - i\beta/2$ and in the second one by $t \to t + i\beta/2$.  As a result
we obtain Eq.~(\ref{eq:dos}).
 
\section{$K(\tau)$ for spinless electrons} 
\label{sec:ksl} 
  
In this appendix we derive the results (\ref{eq:K_0spinless}) and
(\ref{eq:kslresult}) for the correlator $K(\tau)$ in the spinless
case.

\subsection{Evaluation of $K_0(\tau)$,
  Eq.~(\protect\ref{eq:K_0spinless})}
\label{sec:ksl1} 

To derive Eq.~(\ref{eq:K_0spinless}) we evaluate gaussian integral
(\ref{eq:funcint}) under the assumption ${\cal S}'=0$.  First, we find
$\phi_\tau(x,t)$ that minimizes the action ${\cal S}_0+{\cal S}_C(\tau)$.
Differentiating Eqs.~(\ref{eq:S_0}) and (\ref{eq:S_C}) with respect to
$\phi(x,t)$, we find
\[
  \partial_t^2\phi_\tau + v_F^2 \partial_x^2\phi_\tau
  -2v_F E_C\left[n_\tau(t)+\frac{1}{\pi}\phi_\tau-N \right]\delta(x)=0.
\]
The solution of this equation has the form
\begin{equation}
  \label{eq:phi_tau}
  \phi_\tau(x,t)=\pi N -T\sum_{\omega_n}
                 \frac{E_C\exp\left(-\frac{|\omega_n
                 x|}{v_F}\right)}{|\omega_n|+\frac{E_C}{\pi}}
                 n_\tau(\omega_n)e^{-i\omega_n t},
\end{equation}
where $\omega_n=2\pi n T$ are bosonic Matsubara frequencies, and
$n_\tau(\omega_n)$ is the Fourier transform of $n_\tau(t)$,
Eq.~(\ref{eq:ntau}),
\begin{equation} 
\label{eq:nfourier} 
n_\tau(\omega_n)=\frac{e^{i \omega_n \tau}-1}{i\omega_n}.
\end{equation} 

In the calculation of the correlator $K_0(\tau)$ the integrals over the
fluctuations of the field $\phi(x,t)$ about the saddle points
$\phi_\tau(x,t)$ and $\phi_0(x,t)$ in the numerator and the denominator of
Eq.~(\ref{eq:correlator}) cancel each other.  Thus $K_0(\tau)$ is given by
the ratio of the saddle-point values of the respective integrals.
Substituting (\ref{eq:phi_tau}) into (\ref{eq:S_0}) and (\ref{eq:S_C}) we
find the saddle point action in the form
\begin{equation}
  [{\cal S}_0 + {\cal S}_C(\tau)]_{\phi=\phi_\tau(x,t)}=
  \frac{E_C}{\pi^2T}\sum_{n=1}^\infty \frac{1-\cos(2\pi T\tau
  n)}{n\big(n+\frac{E_C}{2\pi^2T}\big)}.
  \label{eq:saddle_point_action}
\end{equation}

In the denominator of Eq.~(\ref{eq:correlator}) we have the saddle point
action at $\tau=0$; according to Eq.~(\ref{eq:saddle_point_action}) it
vanishes.  In the numerator of Eq.~(\ref{eq:correlator}) the time $\tau$
is finite.  Assuming $\tau\gg E_C^{-1}$ and $T\ll E_C$, we find
\begin{equation}
  \label{eq:saddle_point_action_tau}
  -[{\cal S}_0 + {\cal S}_C(\tau)]_{\phi=\phi_\tau(x,t)}
  \simeq 2\ln\frac{\pi^2T}{\gamma E_C|\sin(\pi T\tau)|}.
\end{equation}
The correlator $K_0(\tau)$ is now found by exponentiation of
Eq.~(\ref{eq:saddle_point_action_tau}).  The result is given by
Eq.~(\ref{eq:K_0spinless}).

\subsection{Evaluation of $K(\tau)$ to first order in $r$, 
  Eq.~(\ref{eq:kslresult})}
\label{sec:ksl2}

To derive the first-order correction (\ref{eq:kslresult}) to the
correlator $K_0(\tau)$ one has to evaluate the gaussian functional
integral (\ref{eq:avdef}).  It is convenient to integrate with respect to
fluctuations $\varphi=\phi-\phi_\tau$ of the field $\phi(x,t)$ about the
saddle point $\phi_\tau(x,t)$.  Then the integral (\ref{eq:avdef}) takes
the form
\begin{equation}
  \label{eq:avdef_fluct}
  \langle {\cal S}' \rangle_\tau =  {\rm Re}
         \int_0^\beta dt\, e^{2i\phi_{\tau}(0,t)}
         \left[-\frac{D}{\pi}|r|\left\langle 
           e^{2i\varphi(0,t)}\right\rangle\right],
\end{equation}
where the averaging $\langle\ldots\rangle$ is performed over the
fluctuations around the saddle point $\phi_\tau$.  This averaging can be
viewed as integral (\ref{eq:avdef}) with $n_\tau$ and $N$ in the charging
action ${\cal S}_C$ set to zero.  The evaluation of this integral is
straightforward, but lengthy.  It can be avoided by noticing that the
expression in the square brackets in Eq.~(\ref{eq:avdef_fluct}) is
time-independent and has the meaning of the first-order correction to the
ground state energy of the Hamiltonian
(\ref{eq:freehamiltonian})--(\ref{eq:H_Cspinless}) at $\hat n=N=0$.
Substituting its value found in Ref.~\onlinecite{Matveev95}, we get
\begin{equation}
  \label{eq:avdef_renormalized}
   \langle {\cal S}' \rangle_\tau =  -\frac{\gamma}{\pi^2}|r|E_C\,
         {\rm Re} \int_0^\beta dt\, e^{2i\phi_{\tau}(0,t)}.
\end{equation}
Using Eqs.~(\ref{eq:phi_tau}) and (\ref{eq:nfourier}) we now find
\begin{eqnarray}
  \langle {\cal S}' \rangle_\tau-\langle {\cal S}' \rangle_0 &=&
   -\frac{\gamma}{\pi^2}|r|E_C\,{\rm Re}\,e^{2i\pi N}\nonumber\\
   &&\times\int_0^\beta dt\, \big(e^{i[F(t)-F(t-\tau)]}-1\big),
  \label{eq:avdef_total}
\end{eqnarray}
where
\begin{equation}
  \label{eq:Fdef}
  F(t)=2\sum_{n=1}^\infty \frac{\sin(2\pi Ttn)}{n+\frac{E_C}{2\pi^2T}}.
\end{equation}
At $T\ll E_C$ the series can be evaluated explicitly for arbitrary $t$,
\begin{equation}
  \label{eq:Fapproximate}
  F(t)=
  \begin{cases}{
    \frac{2\pi^2T}{E_C}\cot(\pi Tt), & $t\gg E_C^{-1}$,\cr
    2\int_0^\infty dy\frac{\sin(E_C t y/\pi)}{1+y}, & $t\ll T^{-1}$.}
  \end{cases}
\end{equation}

In order to find the thermoelectric coefficient (\ref{eq:thermocondk}) we
need to find $K(\tau)$ at $\tau\sim T^{-1}\gg E_C^{-1}$.  At these time
scales the details of the short-time behavior of $e^{iF(t)}$ are
irrelevant, and one can replace 
\begin{equation}
  e^{iF(t)}\to 1-\frac{\pi^2\xi}{E_C}\delta(t) + i\sin F(t),
\label{eq:approximation}
\end{equation}
where the constant $\xi\approx1.59$ is defined as
\begin{equation} 
\label{eq:xi} 
\xi= \frac{2}{\pi}  \int_0^\infty dx \left[1-\cos\left(2\int_0^\infty dy 
\frac{\sin(xy)}{1+y} \right) \right].
\end{equation} 
Substituting the approximation (\ref{eq:approximation}) into the integral
in Eq.~(\ref{eq:avdef_total}) and using Eq.~(\ref{eq:Fapproximate}), we
find
\begin{eqnarray}
  \langle {\cal S}' \rangle_\tau-\langle {\cal S}' \rangle_0 &=&
         2\gamma \xi |r|\big[\cos(2\pi N) \nonumber\\
         &&- {\textstyle\frac{2\pi^2T}{E_C}}
                \sin(2\pi N)\cot(\pi T\tau)\big].
  \label{eq:SminusS}
\end{eqnarray}
The calculation of the conductance $G$ and the thermoelectric coefficient
$G_T$ requires the knowledge of the even and odd in $\tau$ components of
$K(\tau)$, respectively.  In Eq.~(\ref{eq:SminusS}) we retained only the
leading-order terms in $T/E_C$ for each of these components.  Substituting
Eq.~(\ref{eq:SminusS}) into (\ref{eq:kptsl}), we arrive at
Eq.~(\ref{eq:kslresult}).

\section{$K(\tau)$ for electrons with spin}
\label{sec:K_spins}

\subsection{Evaluation of $\kappa_c(t,t';\tau)$ and $\kappa_s(t,t')$,
  Eq.~(\ref{eq:kappas})}
\label{sec:kappas}

The correlator $\kappa_c(t,t';\tau)$ defined by Eq.~(\ref{eq:kappa_c}) can
be presented in the form
\begin{equation}
  \label{eq:kappa_c_exp}
  \kappa_c(t,t';\tau)=\frac12 {\rm Re}\,[\kappa_c^+(t,t';\tau)
                                        +\kappa_c^-(t,t';\tau)],
\end{equation}
where
\begin{equation}
  \label{eq:kappas_pm}
  \kappa_c^\pm(t,t';\tau)=\left\langle
          e^{i\sqrt2[\phi_c(0,t)\pm\phi_c(0,t')]}
          \right\rangle_\tau.
\end{equation}
The calculation of the correlators $\kappa_\tau^\pm$ amounts to evaluation
of gaussian integrals.  Similarly to the calculations of
Appendix~\ref{sec:ksl2}, it is convenient to integrate over the
fluctuations $\varphi_c$ about the saddle point
\begin{equation}
  \label{eq:phi_ctau}
  \phi_{c,\tau}(0,t)=\frac{\pi N}{\sqrt2}
                 -T\sum_{\omega_n}
                 \frac{\sqrt{2}E_C}{|\omega_n|+\frac{2E_C}{\pi}}
                 n_\tau(\omega_n)e^{-i\omega_n t},
\end{equation}
where $n(\omega_n)$ is given by Eq.~(\ref{eq:nfourier}).  The saddle point
Eq.~(\ref{eq:phi_ctau}) is easily obtained from (\ref{eq:phi_tau}) by
replacing $E_C\to 2E_C$, $n_\tau(t)\to n_\tau(t)/\sqrt2$, and $N\to
N/\sqrt2$.  Substituting $\phi_c(0,t)=\phi_{c,\tau}(0,t)+\varphi(t)$ into
Eq.~(\ref{eq:kappas_pm}) we find
\begin{eqnarray}
  \kappa_c^\pm(t,t';\tau)&=&
             \exp\left\{i\sqrt2[\phi_{c,\tau}(0,t)
                             \pm\phi_{c,\tau}(0,t')]\right\}\nonumber\\
          &&\times\exp\left\{-2\langle\varphi_c(t)
                   [\varphi_c(t)\pm\varphi_c(t')]\rangle\right\}.
  \label{eq:kappas_pm_saddle}
\end{eqnarray}
To evaluate the last factor in Eq.~(\ref{eq:kappas_pm_saddle}) we 
introduce the generating functional
\begin{equation}
  \label{eq:generating_functional}
  W[\{J(\omega_n)\}]=\left\langle\exp\left[-T\sum_{\omega_n} 
              J(\omega_n) \varphi_c(-\omega_n)\right]\right\rangle.
\end{equation}
This gaussian integral is completely determined by the saddle point value
$\varphi_c^J(t)$ of the field $\varphi_c$,
\[
    W[\{J(\omega_n)\}] =\exp\left[-\frac12 T\sum_{\omega_n} 
              J(\omega_n) \varphi_{c}^J(-\omega_n)\right].
\]
Next we note that fluctuations of $\varphi_c(t)$ coincide with those of
$\phi_c(0,t)$ at $N=n_\tau=0$.  Then $n_\tau(t)$ in
Eq.~(\ref{eq:actionspinb}) plays the role of a source term similar to
$J(t)$.  More precisely, they are related according to
$J(t)=(2\sqrt{2}E_C/\pi) n_\tau(t)$.  Then the saddle point
$\varphi_c^J(t)$ can be determined from (\ref{eq:phi_ctau}) at $N=0$ and
$n_\tau(\omega_n)=(\pi/2\sqrt{2}E_C)J(\omega_n)$, and we obtain
\begin{equation}
  \label{eq:generating_functional_result}
  W[\{J(\omega_n)\}] = \exp\left[\frac{\pi}{4} T\sum_{\omega_n}
                       \frac{J(\omega_n)J(-\omega_n)}
                            {|\omega_n|+\frac{2E_C}{\pi}}\right].
\end{equation}
Differentiating the functional $W$ with respect to $J(\omega_n)$ and
$J(-\omega_m)$, from Eqs.~(\ref{eq:generating_functional}) and
(\ref{eq:generating_functional_result}) we find
\begin{equation}
  \label{eq:fluctuations}
  \langle \varphi_c(-\omega_n) \varphi_c(\omega_m)\rangle
  = \frac{\pi}{2T}\frac{1}{|\omega_n|+\frac{2E_C}{\pi}}\delta_{n,m}.
\end{equation}
In the time representation this result takes the form
\begin{equation}
  \label{eq:time_fluctuations}
  \langle \varphi_c(t) \varphi_c(t')\rangle
    =\frac{\pi}{2}T\sum_{\omega_n}
       \frac{e^{i\omega_n(t-t')}}{|\omega_n|+\frac{2E_C}{\pi}}
       e^{-|\omega_n|/D}.
\end{equation}

The asymptotic behavior of this correlator is
\begin{equation}
  \label{eq:time_fluctuations_asymptotics}
  \langle \varphi_c(t) \varphi_c(t')\rangle =
  \begin{cases}
    {\!\frac12\ln\frac{\pi D}{2\gamma E_C\sqrt{1+[D(t-t')]^2}}, 
      \!& $|t-t'|\ll \frac{1}{E_C}$,\cr
     \!\frac{\pi^4 T^2}{8E_C^2 \sin^2\pi T(t-t')},
      \!& $|t-t'|\gg \frac{1}{E_C}$.}
  \end{cases}
\end{equation}
Substituting Eqs.~(\ref{eq:phi_ctau}), (\ref{eq:kappas_pm_saddle}), and
(\ref{eq:time_fluctuations_asymptotics}), into Eq.~(\ref{eq:kappa_c_exp})
we find the correlator $\kappa_c(t,t';\tau)$ at $|t-t'|\gg E_C^{-1}$ in
the form (\ref{eq:kappa_c_result}), where
\[
    \chi_\tau(t) = \frac{E_C}{2\pi^2 T}\sum_{n=-\infty}^\infty
                 \frac{e^{-i2\pi T(t-\tau) n} - e^{-i2\pi Tt n}}
                      {in\left(|n|+\frac{E_C}{\pi^2T}\right)}.
\]
This definition of $\chi_\tau(t)$ can be rewritten in the form
(\ref{eq:chi}). 

Our derivation of $\kappa_c(t,t';\tau)$ allows one to find
$\kappa_s(t,t')$ as well.  Indeed, at $E_C\to0$ the actions of the charge
and spin modes are identical.  Taking the limit $E_C\to0$ in
Eqs.~(\ref{eq:phi_ctau}) and (\ref{eq:time_fluctuations}), from
(\ref{eq:kappas_pm_saddle}) and (\ref{eq:kappa_c_exp}) we find the
correlator $\kappa_s(t,t')$ at $|t-t'|\gg D^{-1}$ in the form
(\ref{eq:kappa_s_result}).

\subsection{Evaluation of the correlator $K_s^{(1)}(\tau)$,
  Eq.~(\ref{eq:K_s_nonperturbative_result})}

\label{sec:K_s_nonperturbative}

In this Appendix we outline the derivation of the correlator
(\ref{eq:K_s_nonperturbative_result}) starting from
Eq.~(\ref{eq:K_s_first_order}).  At small temperature $T\ll E_C$ one can
expand the expression (\ref{eq:H'_tau}) for $H'_\tau(t)$ to first order in
$\delta\chi_\tau$ and present Eq.~(\ref{eq:K_s_first_order}) in the form
\begin{eqnarray}
  \label{eq:K_s_first_order_Phi}
  K_s^{(1)}(\tau)&=&\Lambda\sin(\pi N)\int_0^\beta  \delta\chi_\tau(t)
                  \Phi(\tau,t)dt,\\
  \label{eq:Phi_definition}
  \Phi(\tau,t)   &=&\langle T_t (c+c^\dagger)_t
                    \Psi(t) U(\tau)U(0)\rangle_0.
\end{eqnarray}
Here we introduced the shorthand notation $(c+c^\dagger)_t\equiv
c(t)+c^\dagger(t)$.  To evaluate $\Phi(\tau,t)$ we substitute the
expression (\ref{eq:unitary}) for $U$.  Since the operator $(c-c^\dagger)$
commutes with the Hamiltonian $H_0$, the Green's function $\langle T_t
(c-c^\dagger)_\tau(c-c^\dagger)_0\rangle_0=-1$, and we find
\begin{equation}
  \label{eq:Phi_no_U}
  \Phi(\tau,t)=\langle T_t (c+c^\dagger)_t
               \Psi(t) (c+c^\dagger)_\tau (c+c^\dagger)_0\rangle_0.
\end{equation}
Considering that the Hamiltonian $H_0$ is quadratic in fermion operators,
one can use Wick's theorem and present $\Phi(\tau,t)$ in terms of the
single particle Green's functions:
\begin{eqnarray}
  \Phi(\tau,t) &=& G_1(\tau)G_2(0) + G_1(t-\tau)G_2(-t) \nonumber\\
               && - G_1(t)G_2(\tau-t).
  \label{eq:Phi_Green}
\end{eqnarray}
Here $G_1(t)$ and $G_2(t)$ are defined as
\begin{mathletters}
\label{eq:Green_functions}
\begin{eqnarray}
  G_1(t)&=&\langle T_t (c+c^\dagger)_t (c+c^\dagger)_0\rangle_0,
  \\
  G_2(t)&=&\langle T_t (c+c^\dagger)_t \Psi(0)\rangle_0.
\end{eqnarray}
\end{mathletters}

Evaluation of the Green's functions (\ref{eq:Green_functions}) can be
facilitated by noticing that upon the substitution $\Lambda\cos(\pi
N)=\lambda$ the Hamiltonian $H_0$ coincides with the Hamiltonian
\begin{equation}
  \label{eq:H_old_fermionized}
  H=\int_{-\infty}^\infty \left[\xi_k c_k^\dagger c_k
         -\lambda(c+c^\dagger)(c_k-c_k^\dagger)\right]dk
\end{equation}
in Eq.~(44) of Ref.~\onlinecite{Matveev95}.  This Hamiltonian was
diagonalized to the form\cite{Matveev95} 
\begin{equation}
  H=E + \int_0^\infty \xi_k
             \left(C_k^\dagger C_k
                   +\tilde C_k^\dagger \tilde C_k\right)dk,
  \label{eq:diagonal}
\end{equation}
where $E$ is the ground state energy of the Hamiltonian $H$ and the fermion
operators $C_k$ and $\tilde C_k$ are given by
\begin{eqnarray}
C_k &=& \frac{\xi_k}{\sqrt{\xi_k^2 + \Gamma^2}}
         \frac{c_k - c_{-k}^\dagger}{\sqrt{2}}
         - \frac{\sqrt{2}\, \lambda}{\sqrt{\xi_k^2 + \Gamma^2}}
           \left(c+c^\dagger\right)
\nonumber\\
       & & +\frac{\Gamma}{\pi\sqrt{\xi_k^2 + \Gamma^2}}
            \int_{-\infty}^{\infty}\frac{d\xi_{k'}}{\xi_k-\xi_{k'}}
            \frac{c_{k'} - c_{-k'}^\dagger}{\sqrt{2}},
  \label{eq:Bogoliubov}\\
\tilde C_k&=&(c_k + c_{-k}^\dagger)/\sqrt{2}.
\end{eqnarray}
Here $\Gamma=4\pi\lambda^2/v_F$, which  in our notations becomes
Eq.~(\ref{eq:Gamma}). 

To find the Green's functions (\ref{eq:Green_functions}) we invert the
transformation (\ref{eq:Bogoliubov}) and obtain
\begin{eqnarray}
  \frac{c_k - c_{-k}^\dagger}{\sqrt{2}} &=& 
            -\frac{\Gamma}{\pi}\int_{-\infty}^{\infty}
                 \frac{d\xi_{k'}}{\xi_k-\xi_{k'}}
            \frac{\theta(\xi_{k'})C_{k'}+\theta(-\xi_{k'})C_{-k'}^\dagger}
                 {\sqrt{\xi_{k'}^2 + \Gamma^2}}\nonumber\\
         &&+\frac{\xi_k}{\sqrt{\xi_k^2 + \Gamma^2}}
            \!\left[\theta(\xi_k)C_k+\theta(-\xi_k)C_{-k}^\dagger\right],
  \\[1ex]
  c+c^\dagger &=& -2^{3/2}\lambda
                \int_0^\infty\frac{dk}
                             {\sqrt{\xi_{k}^2 + \Gamma^2}}
                \left(C_k+C_k^\dagger\right).
\end{eqnarray}
Using these results, the definition of $\Psi$, Eq.~(\ref{eq:Psi}), and the
form (\ref{eq:diagonal}) of the Hamiltonian, we easily obtain the Green's
functions
\begin{eqnarray}
  \label{eq:G_1_result}
  G_1(t) &=& \frac{2\Gamma}{\pi}\,{\rm sgn}\,t
             \int_{-\infty}^\infty\frac{d\xi}{\xi^2 + \Gamma^2}
             \frac{e^{\xi|t|}}{e^{\beta\xi}+1},\\
  G_2(t) &=& -\frac{4\lambda}{v_F}
              \int_{-\infty}^\infty\frac{\xi d\xi}{\xi^2 + \Gamma^2}
              \frac{e^{\xi|t|}}{e^{\beta\xi}+1}.
  \label{eq:G_2_result}
\end{eqnarray}

The Green's functions $G_1(t)$ and $G_2(t)$ are odd and even functions of
$t$, respectively.  Noticing also that $\delta\chi_\tau(t)$ given by
Eq.~(\ref{eq:delta_chi}) is invariant with respect to the change of
variables $t\to \tau-t$, we conclude that the contributions of the second
and third terms in Eq.~(\ref{eq:Phi_Green}) to the integral
(\ref{eq:K_s_first_order_Phi}) are equal to each other.  Finally, the
first term in Eq.~(\ref{eq:Phi_Green}) does not contribute to
(\ref{eq:K_s_first_order_Phi}) because it is independent of $t$, and the
time integral of $\delta\chi_\tau(t)$ vanishes.  Therefore we rewrite
Eq.~(\ref{eq:K_s_first_order_Phi}) as
\begin{equation}
  \label{eq:K_s_Green}
  K_s^{(1)}(\tau)=-2\Lambda\sin(\pi N)\int_0^\beta  \delta\chi_\tau(t)
                     G_1(t)G_2(\tau-t)dt.  
\end{equation}
Without loss of generality we can assume $\Gamma\sim T\ll E_C$.  Since
$\delta\chi_\tau(t)\simeq-(\pi^2T/2E_C) \cot[\pi Tt]$ near $t=0,\beta$, see
Eq.~(\ref{eq:delta_chi_approximate}), the integral (\ref{eq:K_s_Green})
diverges logarithmically at $t\to 0$ and $t\to\beta$.  These divergences
are cut off at the short time scale $E_C^{-1}$ and the long time scale
${\rm min} \{\Gamma^{-1},T^{-1}\}$.  Due to the fact that
$G_1(+0)G_2(\tau)=-G_1(\beta-0)G_2(\tau-\beta)$, the two divergences add
up.  Therefore with logarithmic accuracy the correlator
(\ref{eq:K_s_Green}) is given by
\begin{equation}
  \label{eq:K_s_logarithmic}
  K_s^{(1)}(\tau)=\frac{2\pi\Lambda}{E_C}\sin(\pi N) G_1(+0)G_2(\tau)
                  \ln\frac{E_C}{T+\Gamma}.
\end{equation}
Substituting $G_1(+0)=1$ and the expression (\ref{eq:G_2_result}) for
$G_2$, we arrive at Eq.~(\ref{eq:K_s_nonperturbative_result}).

\end{multicols} 
 
\end{document}